\documentclass[longauth]{aa}      
\usepackage{placeins}
\usepackage{amsmath}
\usepackage{graphicx}
\usepackage{txfonts}
\usepackage{hyperref}
\usepackage{comment}
\usepackage{longtable}
\usepackage{anyfontsize}

\hypersetup{
    colorlinks=true,
    citecolor=blue,
    linkcolor=blue,
    urlcolor=blue,
}

\makeatletter
\renewcommand*\aa@pageof{, page \thepage{} of \pageref*{LastPage}}
\makeatother

\usepackage{color}
\usepackage{xcolor}
\definecolor{byzantine}{rgb}{0.74, 0.2, 0.64}
\definecolor{JungleGreen}{rgb}{0.0, 0.66, 0.61}
\newcommand{\valerio}[1]{\textcolor{black}{#1}}

\newcommand{\trades}[0]{\textsc{TRADES}}
\newcommand{\pyde}[0]{\textsc{PyDE}}
\newcommand{\emcee}[0]{\textsc{emcee}}
\newcommand{\pyorbit}[0]{\textsc{PyORBIT}}

\defcitealias{Petigura2018}{P18}
\defcitealias{Petigura2016}{P16}

\begin{document} 

   \title{The K2-24 planetary system revisited by CHEOPS\thanks{This article uses data from  the CHEOPS program \texttt{CH\_PR100025}. The individual data sets are listed in Table~\ref{tab:log}. Table~\ref{tab:t0}, the light curves and the predicted transit times are only available in electronic form at the CDS: \url{http://cdsweb.u-strasbg.fr/cgi-bin/qcat?J/A+A/}.}}
   \authorrunning{V. Nascimbeni et al.}
   \titlerunning{A new dynamical modeling of K2-24}


   \author{
V. Nascimbeni\thanks{send offprint requests to: \texttt{valerio.nascimbeni@inaf.it}}\inst{1} $^{\href{https://orcid.org/0000-0001-9770-1214}{\includegraphics[scale=0.5]{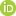}}}$\and 
L. Borsato\inst{1} $^{\href{https://orcid.org/0000-0003-0066-9268}{\includegraphics[scale=0.5]{figures/orcid.jpg}}}$\and 
P. Leonardi\inst{2,1,55} $^{\href{https://orcid.org/0000-0001-6026-9202}{\includegraphics[scale=0.5]{figures/orcid.jpg}}}$\and 
S. G. Sousa\inst{3} $^{\href{https://orcid.org/0000-0001-9047-2965}{\includegraphics[scale=0.5]{figures/orcid.jpg}}}$\and 
T. G. Wilson\inst{4} $^{\href{https://orcid.org/0000-0001-8749-1962}{\includegraphics[scale=0.5]{figures/orcid.jpg}}}$\and 
A. Fortier\inst{5,6} $^{\href{https://orcid.org/0000-0001-8450-3374}{\includegraphics[scale=0.5]{figures/orcid.jpg}}}$\and 
A. Heitzmann\inst{7} $^{\href{https://orcid.org/0000-0002-8091-7526}{\includegraphics[scale=0.5]{figures/orcid.jpg}}}$\and 
G. Mantovan\inst{2,1}\and 
R. Luque\inst{8}\and 
T. Zingales\inst{2,1} $^{\href{https://orcid.org/0000-0001-6880-5356}{\includegraphics[scale=0.5]{figures/orcid.jpg}}}$\and 
G. Piotto\inst{2,1} $^{\href{https://orcid.org/0000-0002-9937-6387}{\includegraphics[scale=0.5]{figures/orcid.jpg}}}$\and 
Y. Alibert\inst{6,5} $^{\href{https://orcid.org/0000-0002-4644-8818}{\includegraphics[scale=0.5]{figures/orcid.jpg}}}$\and 
R. Alonso\inst{10,11} $^{\href{https://orcid.org/0000-0001-8462-8126}{\includegraphics[scale=0.5]{figures/orcid.jpg}}}$\and 
T. Bárczy\inst{12} $^{\href{https://orcid.org/0000-0002-7822-4413}{\includegraphics[scale=0.5]{figures/orcid.jpg}}}$\and 
D. Barrado Navascues\inst{13} $^{\href{https://orcid.org/0000-0002-5971-9242}{\includegraphics[scale=0.5]{figures/orcid.jpg}}}$\and 
S. C. C. Barros\inst{3,14} $^{\href{https://orcid.org/0000-0003-2434-3625}{\includegraphics[scale=0.5]{figures/orcid.jpg}}}$\and 
W. Baumjohann\inst{15} $^{\href{https://orcid.org/0000-0001-6271-0110}{\includegraphics[scale=0.5]{figures/orcid.jpg}}}$\and 
T. Beck\inst{5}\and 
W. Benz\inst{5,6} $^{\href{https://orcid.org/0000-0001-7896-6479}{\includegraphics[scale=0.5]{figures/orcid.jpg}}}$\and 
N. Billot\inst{7} $^{\href{https://orcid.org/0000-0003-3429-3836}{\includegraphics[scale=0.5]{figures/orcid.jpg}}}$\and 
F. Biondi\inst{16,1}\and 
A. Brandeker\inst{17} $^{\href{https://orcid.org/0000-0002-7201-7536}{\includegraphics[scale=0.5]{figures/orcid.jpg}}}$\and 
C. Broeg\inst{5,6} $^{\href{https://orcid.org/0000-0001-5132-2614}{\includegraphics[scale=0.5]{figures/orcid.jpg}}}$\and 
M.-D. Busch\inst{18}\and 
A. Collier Cameron\inst{19} $^{\href{https://orcid.org/0000-0002-8863-7828}{\includegraphics[scale=0.5]{figures/orcid.jpg}}}$\and 
A. C. M. Correia\inst{20}\and 
Sz. Csizmadia\inst{21} $^{\href{https://orcid.org/0000-0001-6803-9698}{\includegraphics[scale=0.5]{figures/orcid.jpg}}}$\and 
P. E. Cubillos\inst{22,15}\and 
M. B. Davies\inst{23} $^{\href{https://orcid.org/0000-0001-6080-1190}{\includegraphics[scale=0.5]{figures/orcid.jpg}}}$\and 
M. Deleuil\inst{24} $^{\href{https://orcid.org/0000-0001-6036-0225}{\includegraphics[scale=0.5]{figures/orcid.jpg}}}$\and 
A. Deline\inst{7}\and 
L. Delrez\inst{25,26,27} $^{\href{https://orcid.org/0000-0001-6108-4808}{\includegraphics[scale=0.5]{figures/orcid.jpg}}}$\and 
O. D. S. Demangeon\inst{3,14} $^{\href{https://orcid.org/0000-0001-7918-0355}{\includegraphics[scale=0.5]{figures/orcid.jpg}}}$\and 
B.-O. Demory\inst{6,5} $^{\href{https://orcid.org/0000-0002-9355-5165}{\includegraphics[scale=0.5]{figures/orcid.jpg}}}$\and 
A. Derekas\inst{28}\and 
B. Edwards\inst{29}\and 
D. Ehrenreich\inst{7,30} $^{\href{https://orcid.org/0000-0001-9704-5405}{\includegraphics[scale=0.5]{figures/orcid.jpg}}}$\and 
A. Erikson\inst{21}\and 
L. Fossati\inst{15} $^{\href{https://orcid.org/0000-0003-4426-9530}{\includegraphics[scale=0.5]{figures/orcid.jpg}}}$\and 
M. Fridlund\inst{31,32} $^{\href{https://orcid.org/0000-0002-0855-8426}{\includegraphics[scale=0.5]{figures/orcid.jpg}}}$\and 
D. Gandolfi\inst{33} $^{\href{https://orcid.org/0000-0001-8627-9628}{\includegraphics[scale=0.5]{figures/orcid.jpg}}}$\and 
K. Gazeas\inst{34} $^{\href{https://orcid.org/0000-0002-8855-3923}{\includegraphics[scale=0.5]{figures/orcid.jpg}}}$\and 
M. Gillon\inst{25} $^{\href{https://orcid.org/0000-0003-1462-7739}{\includegraphics[scale=0.5]{figures/orcid.jpg}}}$\and 
M. Güdel\inst{35}\and 
M. N. Günther\inst{36} $^{\href{https://orcid.org/0000-0002-3164-9086}{\includegraphics[scale=0.5]{figures/orcid.jpg}}}$\and 
Ch. Helling\inst{15,37}\and 
K. G. Isaak\inst{36} $^{\href{https://orcid.org/0000-0001-8585-1717}{\includegraphics[scale=0.5]{figures/orcid.jpg}}}$\and 
F. Kerschbaum\inst{35}\and 
L. L. Kiss\inst{38,39}\and 
J. Korth\inst{40} $^{\href{https://orcid.org/0000-0002-0076-6239}{\includegraphics[scale=0.5]{figures/orcid.jpg}}}$\and 
K. W. F. Lam\inst{21} $^{\href{https://orcid.org/0000-0002-9910-6088}{\includegraphics[scale=0.5]{figures/orcid.jpg}}}$\and 
J. Laskar\inst{41} $^{\href{https://orcid.org/0000-0003-2634-789X}{\includegraphics[scale=0.5]{figures/orcid.jpg}}}$\and 
A. Lecavelier des Etangs\inst{42} $^{\href{https://orcid.org/0000-0002-5637-5253}{\includegraphics[scale=0.5]{figures/orcid.jpg}}}$\and 
A. Leleu\inst{7,5} $^{\href{https://orcid.org/0000-0003-2051-7974}{\includegraphics[scale=0.5]{figures/orcid.jpg}}}$\and 
M. Lendl\inst{7} $^{\href{https://orcid.org/0000-0001-9699-1459}{\includegraphics[scale=0.5]{figures/orcid.jpg}}}$\and 
D. Magrin\inst{1} $^{\href{https://orcid.org/0000-0003-0312-313X}{\includegraphics[scale=0.5]{figures/orcid.jpg}}}$\and 
P. F. L. Maxted\inst{43} $^{\href{https://orcid.org/0000-0003-3794-1317}{\includegraphics[scale=0.5]{figures/orcid.jpg}}}$\and 
B. Merín\inst{44} $^{\href{https://orcid.org/0000-0002-8555-3012}{\includegraphics[scale=0.5]{figures/orcid.jpg}}}$\and 
C. Mordasini\inst{5,6}\and 
G. Olofsson\inst{17} $^{\href{https://orcid.org/0000-0003-3747-7120}{\includegraphics[scale=0.5]{figures/orcid.jpg}}}$\and 
R. Ottensamer\inst{35}\and 
I. Pagano\inst{45} $^{\href{https://orcid.org/0000-0001-9573-4928}{\includegraphics[scale=0.5]{figures/orcid.jpg}}}$\and 
E. Pallé\inst{10,11} $^{\href{https://orcid.org/0000-0003-0987-1593}{\includegraphics[scale=0.5]{figures/orcid.jpg}}}$\and 
G. Peter\inst{46} $^{\href{https://orcid.org/0000-0001-6101-2513}{\includegraphics[scale=0.5]{figures/orcid.jpg}}}$\and 
D. Pollacco\inst{4}\and 
D. Queloz\inst{47,48} $^{\href{https://orcid.org/0000-0002-3012-0316}{\includegraphics[scale=0.5]{figures/orcid.jpg}}}$\and 
R. Ragazzoni\inst{1,9} $^{\href{https://orcid.org/0000-0002-7697-5555}{\includegraphics[scale=0.5]{figures/orcid.jpg}}}$\and 
N. Rando\inst{36}\and 
H. Rauer\inst{21,49} $^{\href{https://orcid.org/0000-0002-6510-1828}{\includegraphics[scale=0.5]{figures/orcid.jpg}}}$\and 
I. Ribas\inst{50,51} $^{\href{https://orcid.org/0000-0002-6689-0312}{\includegraphics[scale=0.5]{figures/orcid.jpg}}}$\and 
N. C. Santos\inst{3,14} $^{\href{https://orcid.org/0000-0003-4422-2919}{\includegraphics[scale=0.5]{figures/orcid.jpg}}}$\and 
G. Scandariato\inst{45} $^{\href{https://orcid.org/0000-0003-2029-0626}{\includegraphics[scale=0.5]{figures/orcid.jpg}}}$\and 
D. Ségransan\inst{7} $^{\href{https://orcid.org/0000-0003-2355-8034}{\includegraphics[scale=0.5]{figures/orcid.jpg}}}$\and 
A. E. Simon\inst{5,6} $^{\href{https://orcid.org/0000-0001-9773-2600}{\includegraphics[scale=0.5]{figures/orcid.jpg}}}$\and 
A. M. S. Smith\inst{21} $^{\href{https://orcid.org/0000-0002-2386-4341}{\includegraphics[scale=0.5]{figures/orcid.jpg}}}$\and 
R. Southworth\inst{52}\and 
M. Stalport\inst{26,25}\and 
S. Sulis\inst{24} $^{\href{https://orcid.org/0000-0001-8783-526X}{\includegraphics[scale=0.5]{figures/orcid.jpg}}}$\and 
Gy. M. Szabó\inst{28,53} $^{\href{https://orcid.org/0000-0002-0606-7930}{\includegraphics[scale=0.5]{figures/orcid.jpg}}}$\and 
S. Udry\inst{7} $^{\href{https://orcid.org/0000-0001-7576-6236}{\includegraphics[scale=0.5]{figures/orcid.jpg}}}$\and 
B. Ulmer\inst{46}\and 
V. Van Grootel\inst{26} $^{\href{https://orcid.org/0000-0003-2144-4316}{\includegraphics[scale=0.5]{figures/orcid.jpg}}}$\and 
J. Venturini\inst{7} $^{\href{https://orcid.org/0000-0001-9527-2903}{\includegraphics[scale=0.5]{figures/orcid.jpg}}}$\and 
E. Villaver\inst{10,11}\and 
N. A. Walton\inst{54} $^{\href{https://orcid.org/0000-0003-3983-8778}{\includegraphics[scale=0.5]{figures/orcid.jpg}}}$
          }

   \institute{
INAF, Osservatorio Astronomico di Padova, Vicolo dell'Osservatorio 5, 35122 Padova, Italy \and
Dipartimento di Fisica e Astronomia, Università degli Studi di Padova, Vicolo dell’Osservatorio 3, 35122 Padova, Italy \and
Instituto de Astrofisica e Ciencias do Espaco, Universidade do Porto, CAUP, Rua das Estrelas, 4150-762 Porto, Portugal \and
Department of Physics, University of Warwick, Gibbet Hill Road, Coventry CV4 7AL, United Kingdom \and
Weltraumforschung und Planetologie, Physikalisches Institut, University of Bern, Gesellschaftsstrasse 6, 3012 Bern, Switzerland \and
Center for Space and Habitability, University of Bern, Gesellschaftsstrasse 6, 3012 Bern, Switzerland \and
Observatoire astronomique de l'Université de Genève, Chemin Pegasi 51, 1290 Versoix, Switzerland \and
Department of Astronomy \& Astrophysics, University of Chicago, Chicago, IL 60637, USA \and
Dipartimento di Fisica e Astronomia "Galileo Galilei", Università degli Studi di Padova, Vicolo dell'Osservatorio 3, 35122 Padova, Italy \and
Instituto de Astrofísica de Canarias, Vía Láctea s/n, 38200 La Laguna, Tenerife, Spain \and
Departamento de Astrofísica, Universidad de La Laguna, Astrofísico Francisco Sanchez s/n, 38206 La Laguna, Tenerife, Spain \and
Admatis, 5. Kandó Kálmán Street, 3534 Miskolc, Hungary \and
Depto. de Astrofísica, Centro de Astrobiología (CSIC-INTA), ESAC campus, 28692 Villanueva de la Cañada (Madrid), Spain \and
Departamento de Fisica e Astronomia, Faculdade de Ciencias, Universidade do Porto, Rua do Campo Alegre, 4169-007 Porto, Portugal \and
Space Research Institute, Austrian Academy of Sciences, Schmiedlstrasse 6, A-8042 Graz, Austria \and
Max Planck Institute for Extraterrestrial Physics, Gießenbachstraße, 85748 Garching, Germany \and
Department of Astronomy, Stockholm University, AlbaNova University Center, 10691 Stockholm, Sweden \and
Physikalisches Institut, University of Bern, Gesellschaftsstrasse 6, 3012 Bern, Switzerland \and
Centre for Exoplanet Science, SUPA School of Physics and Astronomy, University of St Andrews, North Haugh, St Andrews KY16 9SS, UK \and
CFisUC, Departamento de F\'isica, Universidade de Coimbra, 3004-516 Coimbra, Portugal \and
Institute of Planetary Research, German Aerospace Center (DLR), Rutherfordstrasse 2, 12489 Berlin, Germany \and
INAF, Osservatorio Astrofisico di Torino, Via Osservatorio, 20, I-10025 Pino Torinese To, Italy \and
Centre for Mathematical Sciences, Lund University, Box 118, 221 00 Lund, Sweden \and
Aix Marseille Univ, CNRS, CNES, LAM, 38 rue Frédéric Joliot-Curie, 13388 Marseille, France \and
Astrobiology Research Unit, Université de Liège, Allée du 6 Août 19C, B-4000 Liège, Belgium \and
Space sciences, Technologies and Astrophysics Research (STAR) Institute, Université de Liège, Allée du 6 Août 19C, 4000 Liège, Belgium \and
Institute of Astronomy, KU Leuven, Celestijnenlaan 200D, 3001 Leuven, Belgium \and
ELTE Gothard Astrophysical Observatory, 9700 Szombathely, Szent Imre h. u. 112, Hungary \and
SRON Netherlands Institute for Space Research, Niels Bohrweg 4, 2333 CA Leiden, Netherlands \and
Centre Vie dans l’Univers, Faculté des sciences, Université de Genève, Quai Ernest-Ansermet 30, 1211 Genève 4, Switzerland \and
Leiden Observatory, University of Leiden, PO Box 9513, 2300 RA Leiden, The Netherlands \and
Department of Space, Earth and Environment, Chalmers University of Technology, Onsala Space Observatory, 439 92 Onsala, Sweden \and
Dipartimento di Fisica, Università degli Studi di Torino, via Pietro Giuria 1, I-10125, Torino, Italy \and
National and Kapodistrian University of Athens, Department of Physics, University Campus, Zografos GR-157 84, Athens, Greece \and
Department of Astrophysics, University of Vienna, Türkenschanzstrasse 17, 1180 Vienna, Austria \and
European Space Agency (ESA), European Space Research and Technology Centre (ESTEC), Keplerlaan 1, 2201 AZ Noordwijk, The Netherlands \and
Institute for Theoretical Physics and Computational Physics, Graz University of Technology, Petersgasse 16, 8010 Graz, Austria \and
Konkoly Observatory, Research Centre for Astronomy and Earth Sciences, 1121 Budapest, Konkoly Thege Miklós út 15-17, Hungary \and
ELTE E\"otv\"os Lor\'and University, Institute of Physics, P\'azm\'any P\'eter s\'et\'any 1/A, 1117 Budapest, Hungary \and
Lund Observatory, Division of Astrophysics, Department of Physics, Lund University, Box 118, 22100 Lund, Sweden \and
IMCCE, UMR8028 CNRS, Observatoire de Paris, PSL Univ., Sorbonne Univ., 77 av. Denfert-Rochereau, 75014 Paris, France \and
Institut d'astrophysique de Paris, UMR7095 CNRS, Université Pierre \& Marie Curie, 98bis blvd. Arago, 75014 Paris, France \and
Astrophysics Group, Lennard Jones Building, Keele University, Staffordshire, ST5 5BG, United Kingdom \and
European Space Agency, ESA - European Space Astronomy Centre, Camino Bajo del Castillo s/n, 28692 Villanueva de la Cañada, Madrid, Spain \and
INAF, Osservatorio Astrofisico di Catania, Via S. Sofia 78, 95123 Catania, Italy \and
Institute of Optical Sensor Systems, German Aerospace Center (DLR), Rutherfordstrasse 2, 12489 Berlin, Germany \and
ETH Zurich, Department of Physics, Wolfgang-Pauli-Strasse 2, CH-8093 Zurich, Switzerland \and
Cavendish Laboratory, JJ Thomson Avenue, Cambridge CB3 0HE, UK \and
Institut fuer Geologische Wissenschaften, Freie Universitaet Berlin, Maltheserstrasse 74-100,12249 Berlin, Germany \and
Institut de Ciencies de l'Espai (ICE, CSIC), Campus UAB, Can Magrans s/n, 08193 Bellaterra, Spain \and
Institut d'Estudis Espacials de Catalunya (IEEC), 08860 Castelldefels (Barcelona), Spain \and
ESOC, European Space Agency, Robert-Bosch-Str. 5, 64293 Darmstadt, Germany \and
HUN-REN-ELTE Exoplanet Research Group, Szent Imre h. u. 112., Szombathely, H-9700, Hungary \and
Institute of Astronomy, University of Cambridge, Madingley Road, Cambridge, CB3 0HA, United Kingdom \and
Dipartimento di Fisica,  Università di Trento, Via Sommarive 14, 38123 Povo (TN), Italy
}
   \date{Submitted 23 May 2024 / Accepted 4 September 2024}

  \abstract{The planetary system K2-24 is composed of two transiting low-density Neptunians locked in an almost perfect 2:1 resonance and showing large transit time variations (TTVs), and it is an excellent laboratory to search for signatures of planetary migration. Previous studies performed with K2, Spitzer, and RV data tentatively claimed a significant non-zero eccentricity for one or both planets, possibly high enough to challenge the scenario of pure disk migration through resonant capture. With 13 new CHEOPS light curves (seven of planet b, six of planet c), we carried out a global photometric and dynamical re-analysis by including all the available literature data as well. We obtained the most accurate set of planetary parameters to date for the K2-24 system, including radii and masses at 1\% and 5\% precision (now essentially limited by the uncertainty on stellar parameters) and non-zero eccentricities $e_b=0.0498_{-0.0018}^{+0.0011}$, $e_c=0.0282_{-0.0007}^{+0.0003}$ detected at very high significance for both planets. Such relatively large values imply the need for an additional physical mechanism of eccentricity excitation during or after the migration stage. Also, while the accuracy of the previous TTV model had drifted by up to 0.5 days at the current time, we constrained the orbital solution firmly enough to predict the forthcoming transits for the next $\sim$15 years, thus enabling efficient follow-up with top-level facilities such as JWST or ESPRESSO.
  }


   \maketitle
%
\section{Introduction}\label{sec:introduction}

The advent of space-based high-precision photometry, inaugurated by CoRoT \citep{Auvergne2009} and then continued by \emph{Kepler} \citep{Borucki2010}, its second phase K2 \citep{Howell2014}, and more recently by TESS \citep{Ricker2015}, has enabled not only the unexpected discovery of entirely new classes of exoplanets but also the application of analysis techniques hitherto relegated to theory. Among these, one of the most fruitful is a dynamical technique known as transit time variations (TTVs; \citealt{Agol2005}), in which the gravitational perturbation between planets and its time-variable effect on the measured orbital periods is exploited to retrieve their orbital solution. In the case where two or more planets transit their host stars, TTVs are very effective at both confirming the planetary nature of candidates and measuring their masses without the need of (or in synergy with; \citealt{Malavolta2017}) radial velocity (RV) measurements.

Dynamical simulations have shown that the expected amplitude of TTVs in ordinary planetary systems is quite small, usually on the order of magnitude of seconds to minutes \citep{Holman2005}. Such an amplitude is close or even below the detection limit imposed by photon noise and/or stellar activity \citep{Barros2013}. A fairly interesting exception is represented by systems where planets are locked in mean-motion resonances (MMRs) or, more generally, close to commensurability,\footnote{Being close to an integer ratio of orbital periods does not necessarily imply the system is in an MMR from a dynamical point of view; see also our discussion in Section~\ref{sec:mmr}.} the orbital period ratio being close to an integer ratio. Low-order MMRs in the $j+1\!\!:\!j$ form ($j\in\mathbb{N}$), such as 2:1 or 3:2 can boost the TTV signal by orders of magnitude, reaching hours or even days in the most favorable configurations \citep{Agol2018}. A famous case, and the first one to be investigated, is the \emph{Kepler}-9 system \citep{Holman2010}, a pair of transiting warm Saturn-sized planets orbiting their host in about 19.2 and 38.9 days, that is, close to a 2:1 MMR. Resonant configurations are not merely a useful playground to exploit the TTV technique. Rather, they are also extremely interesting by themselves since they represent a unique laboratory to test planetary formation and migration theories \citep{Batygin2015}. In particular, how resonances can be maintained during a disk-migration phase and form or change at a later stage is currently a very active area of debate (\citealt{Huang2023} and references therein).

The typical timescale of the orbital period modulation induced by the TTV (sometimes called the ``superperiod''; $P_\mathrm{TTV}$) needs to be fully mapped to avoid degeneracy in the dynamical retrieval, and it can reach months or even years. In the \citet{Hadden2016} approximation, the superperiod can be estimated as a function of the orbital periods of the inner and outer planets:
\begin{equation}
    P_\mathrm{ttv} = \left | \frac{j+1}{P_\mathrm{out}}-\frac{j}{P_\mathrm{in}}\right |^{-1} \textrm{ .}
\end{equation}
For most orbital configurations, $P_\mathrm{TTV}$ can be significantly longer than the average duration of a K2 campaign ($\sim$ 70-80~days) or a TESS sector ($\sim$ 27~days). Even if TESS, by design, is able to revisit a given target throughout additional sectors (according mostly to its ecliptic latitude), the sampling of the transit times can be very sparse, especially in the long-period regime ($P\gtrsim$ 30~d). This is particularly true for those systems discovered by K2, which lie close to the ecliptic and are therefore rarely monitored by TESS, if at all. 

On the other hand, the ESA S-class mission CHEOPS \citep{Benz2021,Fortier2024}, launched in 2019, is very efficient at observing at low ecliptic latitudes, and being a single-target telescope, it has the ability to gather even very long-period transits, once their ephemeris is reasonably constrained. CHEOPS has been successfully exploited several times to follow up on systems discovered by K2, sometimes with a particular focus on TTV analysis. One such system is WASP-47 \citep{Nascimbeni2023}, for which our analysis led to an improvement of the orbital and physical parameters and, in particular, of the density of planet d. We chose K2-24 as the next system to explore for the science case mentioned above within the CHEOPS GTO program.

The system K2-24, announced by \citet{Petigura2016} (hereafter P16), is a planetary system made by two (sub-)Saturn-sized ($R_b\simeq 5.6$~$R_\oplus$, $R_c\simeq 8$~$R_\oplus$) planets close to a 2:1 period ratio, with orbital periods of $P_b\simeq 21$ and $P_c\simeq 42$~days. Since the baseline of the K2 light curve was not long enough to detect TTVs, the discovery paper had to rely on the HIRES RVs alone to constrain the planetary masses, which turned out to be in the Neptunian range ($M_b= 21.0\pm 5.4$~$M_\oplus$, $M_c= 27\pm 7$~$M_\oplus$), hence making K2-24b and -c extremely inflated planets with unusually large H/He envelopes predicted by models. A further PSF/HARPS RV follow-up by \citet{Dai2016} essentially confirmed the mass estimates at $M_b= 20\pm 4$~$M_\oplus$ and $M_c= 26\pm 6$~$M_\oplus$. The only follow-up transits published so far (two of  -b and two of -c), observed by Spitzer in 2015-2016, were presented by \citet{Petigura2018} (hereafter P18), who also merged all the existing photometric and spectroscopic data and carried out the first TTV analysis of this system, revealing even smaller masses ($M_b= 19\pm 2$~$M_\oplus$, $M_c= 15\pm 2$~$M_\oplus$) and tentatively detecting the presence of an outer $54\pm14$~$M_\oplus$ companion at $\sim 1.1$~au. 

The TTV modeling by \citetalias{Petigura2018}, performed through the analytic approach developed by \citet{Lithwick2012} and based on transit data covering only $\sim40$\% of $P_\mathrm{TTV}$, did not yield a precise measurement for both eccentricities $e_b$ and $e_c$. Rather, it concluded that at least one planet must have an eccentricity significantly larger than zero, adopting $e_b=0.06\pm 0.01$ and $e_c<0.07$ (at 90\% confidence) based on dynamical stability constraints and an imposed prior derived from the distribution of $\langle e \rangle$ observed in \emph{Kepler} multi-planet systems. \citet{Antoniadou2020} later presented a more detailed analysis of dynamical stability in K2-24 based on the planetary parameters by \citetalias{Petigura2018}, concluding that MMR locking protects its long-term evolution and more tightly constraining the eccentricity of the outer planet to $e_c<0.05$.

\citet{Teyssandier2020} further investigated the dynamical architecture of K2-24 and its implications for its formation and migration history, concluding that a pure disc-induced migration is not able to reproduce the period ratio and the TTV amplitude observed and that it would result in much smaller eccentricities, by a factor of $\sim 30$. Rather, they proposed a two-stage scenario where the two planets are first captured in resonance at low eccentricities within the disk. Then eccentricities are excited by an outer companion (such as the one hinted at by RV observations) during the disk dispersal phase. The same authors also suggested that the actual value of $e_b$ and $e_c$ may be higher than the \citetalias{Petigura2018} estimate, according to their simulations.

\begin{figure}
    \centering
    \includegraphics[width=\columnwidth]{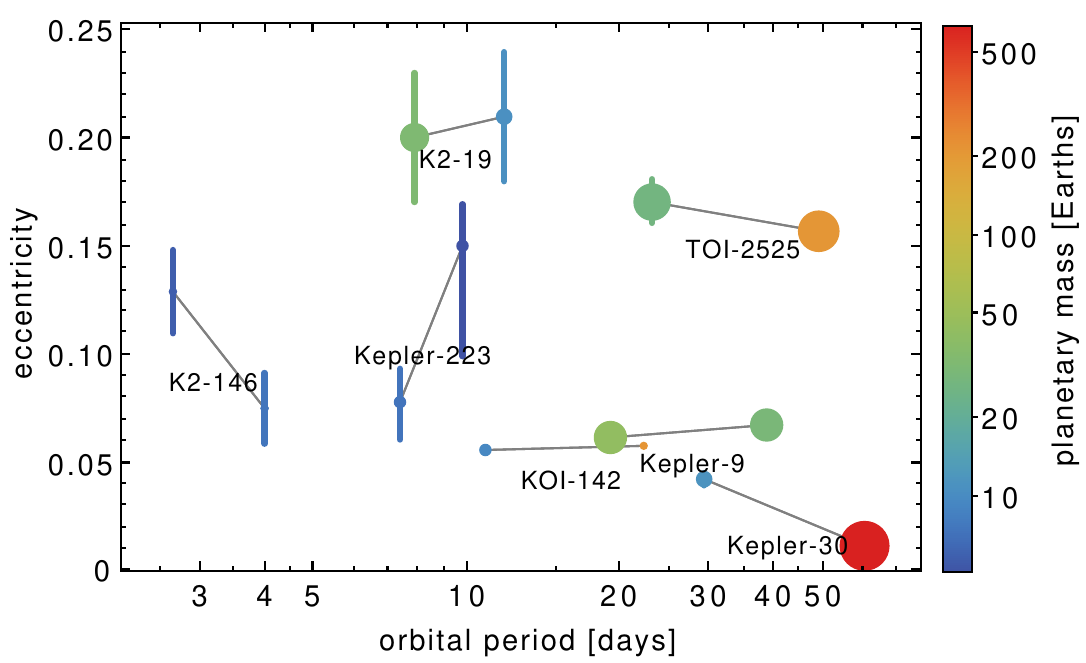}
    \caption{Orbital eccentricities constrained at better than $3\sigma$ for resonant pairs of transiting planets, from the NASA Exoplanet Archive (see text for references). Each pair is linked with a gray segment and labelled with the host star name. The point size is proportional to the planetary radius, while the planetary mass is color-coded.}
    \label{fig:mmr_ecc}
\end{figure}

Only for a handful of transiting planetary systems are there accurate eccentricities for planets in or close to low-order MMRs. From the latest version of the NASA Exoplanet Archive (\citealt{Akeson2013}; v. 2023-12-28), only seven\footnote{\valerio{An additional eighth system is TIC279401253 \citep{Bozhilov2023}, a 2:1 pair of giants ($P_b\simeq 77$~d, $P_c\simeq 155$~d) whose outer one is not transiting and detected through RVs.}} pairs of resonant\footnote{All the listed pairs lie in or close to the 3:2 MMR (K2-146, \emph{Kepler}-223, K2-19) or the 2:1 MMR (all the others).}  planets can be found with both eccentricities constrained at better than 3$\sigma$ (Fig.~\ref{fig:mmr_ecc}). Sorted by increasing average orbital period, they are K2-146 b/c \citep{Hamann2019}, \emph{Kepler}-223 b/c \citep{Mills2016}, K2-19 b/c \citep{Petigura2020}, KOI-142=\emph{Kepler}-88 b/c \citep{Weiss2020}, \emph{Kepler}-9 b/c \citep{Borsato2019}, TOI-2525 b/c \citep{Trifonov2023}, and \emph{Kepler}-30 b/c \citep{SanchisOjeda2012}. Among these, only K2-146 and \emph{Kepler}-223 lie in the Neptunian mass regime but are quite close-in at $P<10$~d, where tidal interactions with the host star start to become significant \citep{LithwickWu2012}. From this point of view, the K2-24 system offers us the rare opportunity to probe the ``primordial'' eccentricities of a pair of warm ($T_\mathrm{eq}<800$~K) Neptunians (i.e., not affected by tidal effects).

The aim of this paper is to present and analyze thirteen new CHEOPS observations of K2-24-b/c and to merge them with all the existing literature light curves and RVs in order to derive an updated and consistent dynamical solution able to 1) recover the transit ephemeris for any future follow-up; 2) improve the measurement of planetary masses, radii, and densities, with a particular focus on the implications for their inner structure; and 3) firmly detect eccentricities for both planets. We present the new observations together with the employed archival data in Section~\ref{sec:observations}. We describe the photometric modeling of the light curves in Section~\ref{sec:photometric} and then the global TTV+RV dynamical analysis in Section~\ref{sec:dynamical}. Finally, the results are compared and interpreted in Section~\ref{sec:conclusions}, where prospects for the future characterization of this system are also discussed.

\section{Observations}\label{sec:observations}

We collected all the available photometric and spectroscopic data of K2-24 for our analysis; they are described in the following subsections.  The very long orbital periods of K2-24b and -c, together with their long duration, uncertain ephemeris (See Section~\ref{sec:observations_cheops}), and small transit depths (2-4~mmag), make the ground-based follow-up of this system extremely difficult. Indeed, no ground-based light curves have been published so far. It is also worth mentioning that K2-24 was not observed by TESS in its first six observing cycles (2018-2024). All the time stamps of the photometric and spectroscopic data described below were uniformly converted to the BJD-TDB standard and referred to the mid-exposure instant, following the prescription by \citet{Eastman2010}.

\subsection{K2 and Spitzer photometry}\label{sec:observations_k2}

K2-24 has been observed by K2 once, in Campaign 2, from 2014-08-23 to 2014-11-13. This uninterrupted, $\sim75$~day-long light curve contains four transits of planet -b and two transits of planet -c (plotted with green points in Fig.~\ref{fig:lck2hst}) and led to their discovery published by \citetalias{Petigura2016}. This light curve has been corrected for systematic errors due to the spacecraft jitter and drifting by following the approach developed by \citet{VanderburgJohnson2014}.

Four more transits were secured by Spitzer and presented by \citetalias{Petigura2018}: two of planet -b on 2015-10-27 and 2016-06-13, and two of planet -c on 2015-11-12 and 2016-06-10. Both of the time series of planet c actually cover partial transits. The scheduling was based on a simple linear ephemeris since a more sophisticated TTV model was not yet available at that time. All the Spitzer light curves have been corrected for systematics through the pixel-level decorrelation algorithm (PLD; \citealt{Deming2015}) as modified by \citet{Benneke2017}.

\subsection{HST photometry}\label{sec:observations_hst}
We downloaded the publicly available HST WFC3 G141 observations of K2-24b from the MAST archive. These data cover a single transit gathered on 2016-07-04 as part of proposal GO-14455 (PI: E.~Petigura, plotted with orange points in Fig.~\ref{fig:lck2hst}) to extract a transmission spectrum of the planetary atmosphere. These observations have been analyzed previously by \citet{Edwards2023b} -- we return to their results in Section~\ref{sec:conclusions}.
The visit consists of a total of eight HST orbits. In our analysis, we excluded the first one due to the presence of significant time-dependent systematic errors. At the beginning of each orbit, a direct image captured with the F130N filter was used for wavelength calibration. These data were collected with the GRISM256 aperture and the SPARS10 reading sequence. The total exposure time was set to 103.13~s, with 16 up-the-ramp reads for exposure. Both scanning directions were employed. 

We calibrated the raw WFC3 data and extracted the photometric information through the \textsc{Iraclis} dedicated pipeline (\citealt{Tsiaras2016a}, \citeyear{Tsiaras2016b}, \citeyear{Tsiaras2018}). We extracted the detrended white-light curve (spectral range: 1.088 to 1.680 $\mu$m; plotted with orange points in Fig.~\ref{fig:lck2hst}) from the calibrated images, taking into account the tilted configuration of the WFC3/NIR detector and modeling the time-dependent systematics using the Eq.~1 of \citet{Edwards2023a}. The HST WFC3 time series are often affected by linear long-term and exponential short-term (``orbit ramps'') trends, especially when observing bright sources. We note that both ingress and egress are missing from this visit, implying that the transit time $T_0$ is expected to be relatively poorly constrained.

\subsection{CHEOPS photometry}\label{sec:observations_cheops}

\begin{figure*}
    \centering
    \includegraphics[width=0.99\columnwidth]{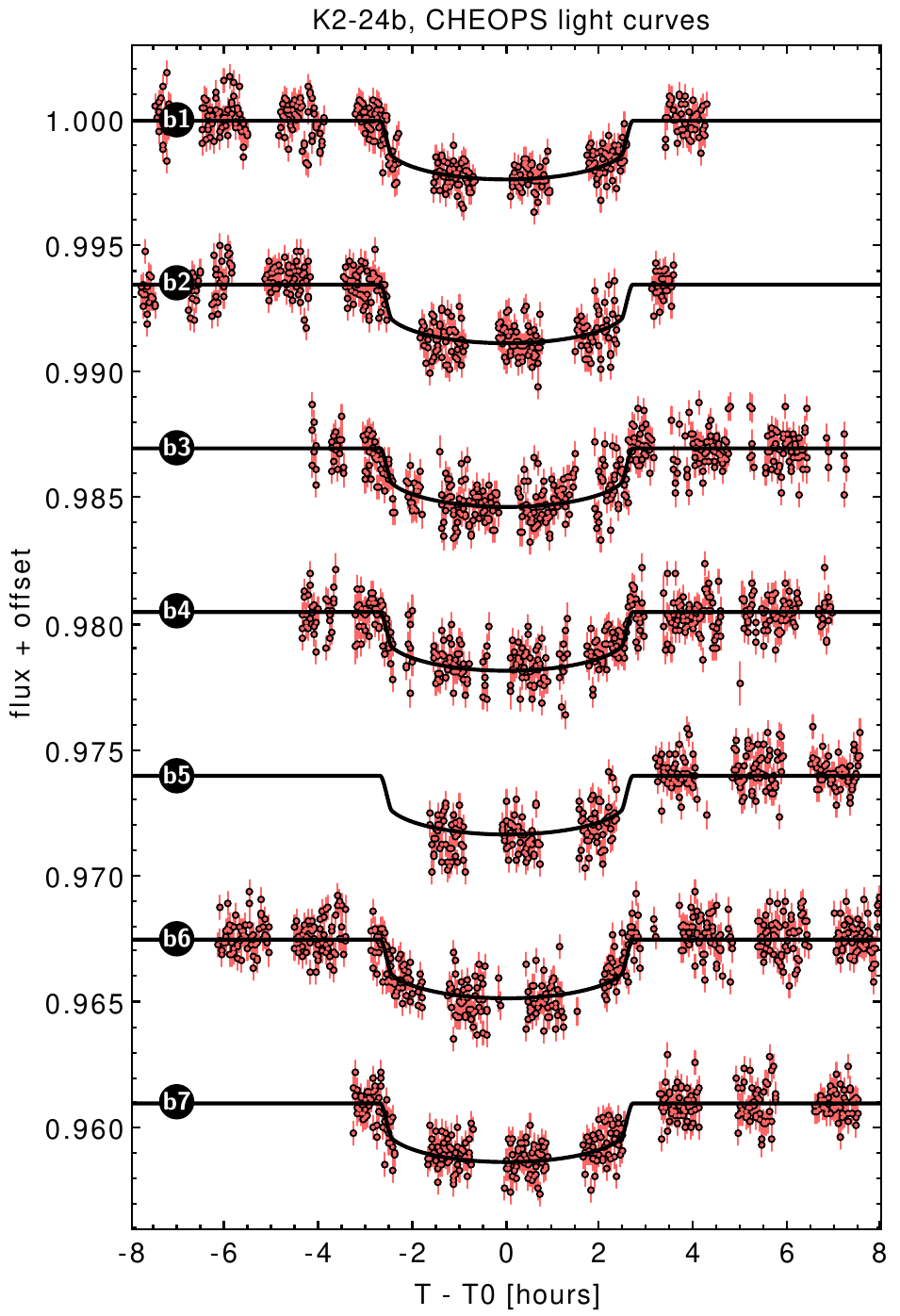}\hspace{0.01\columnwidth}
    \includegraphics[width=0.99\columnwidth]{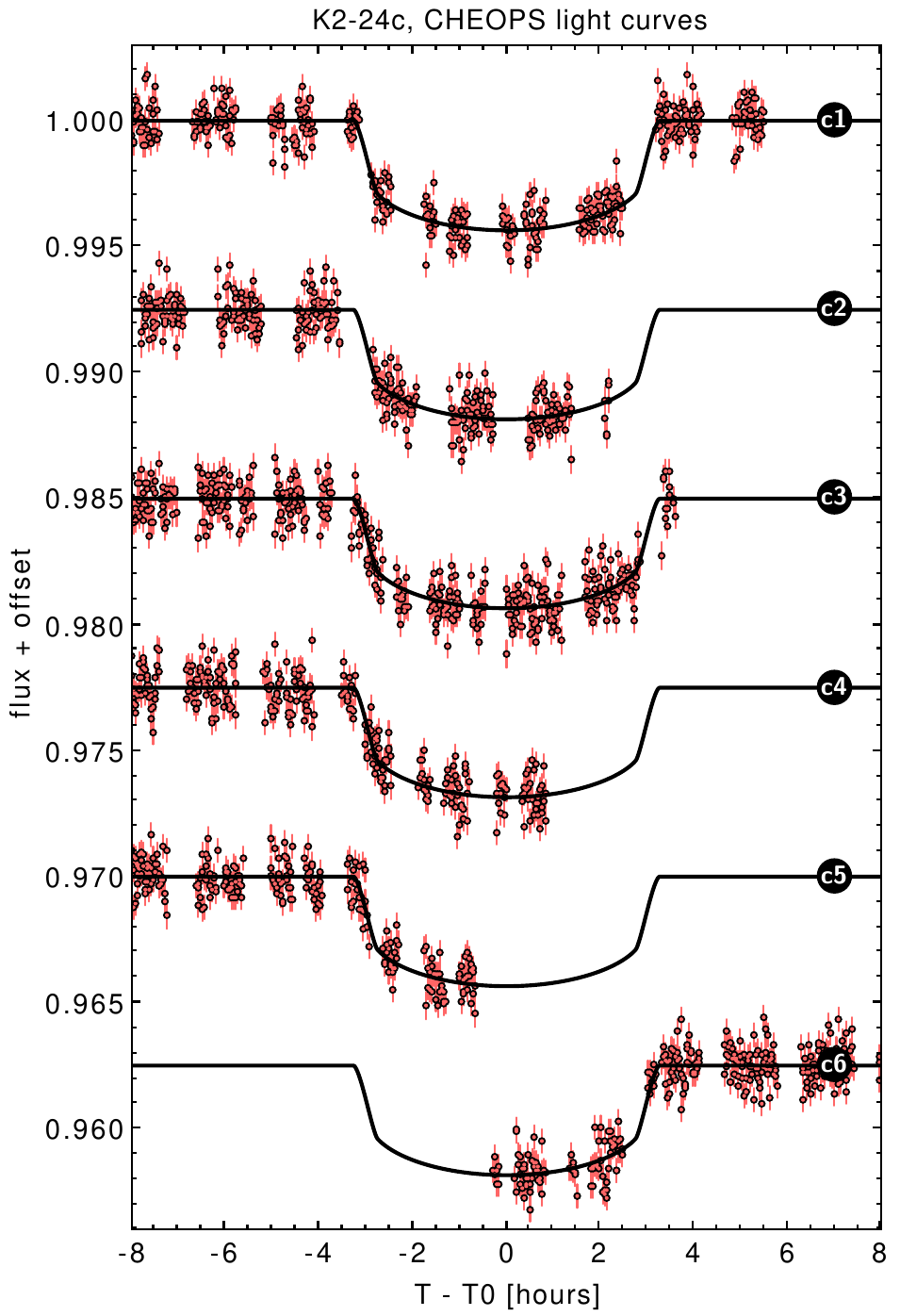}\\
    \caption{Light curves of K2-24b (left panel) and K2-24c (right panel) from CHEOPS analyzed for the present work, after detrending. For each light curve, the corresponding label matches the ID given in Table~\ref{tab:log}. Arbitrary vertical offsets of 0.065 and 0.0075 were added, respectively, to both sets for visualization purposes.}
    \label{fig:lc}
\end{figure*}

K2-24 was targeted by CHEOPS thirteen times over a span of about two years within the GTO subprogram \#25 (PID~\texttt{PR100025}) focused on the study of the mass-radius relation through the TTV analysis of resonant pairs of low-mass exoplanets. A complete log of the observations is reported in Table \ref{tab:log}. The corresponding light curves, extracted by the CHEOPS DRP v14 pipeline \citep{Hoyer2020}, are plotted in Fig.~\ref{fig:lc} and labelled with matching IDs. The gaps located at regular time intervals are due to the avoidance angles of CHEOPS and to the SAA crossing events, Earth occultations, and Earth stray light contamination during its 98.77-min low Earth orbit.

\begin{figure*}
    \centering
    \includegraphics[width=0.99\columnwidth]{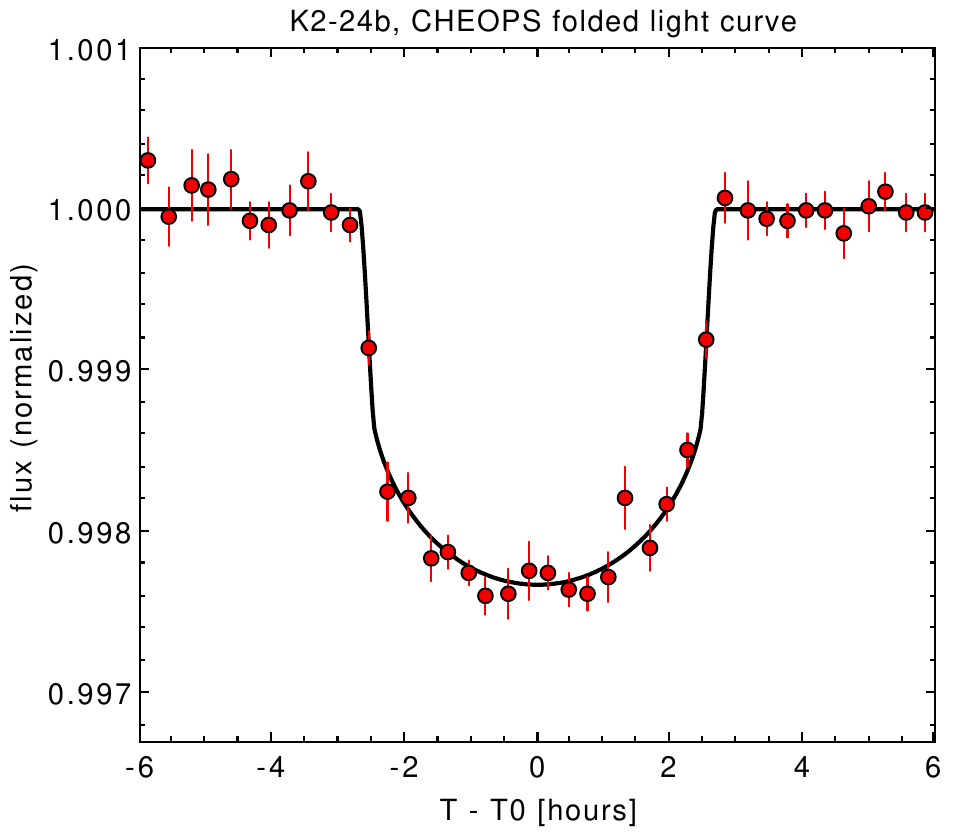}\hspace{0.01\columnwidth}
    \includegraphics[width=0.99\columnwidth]{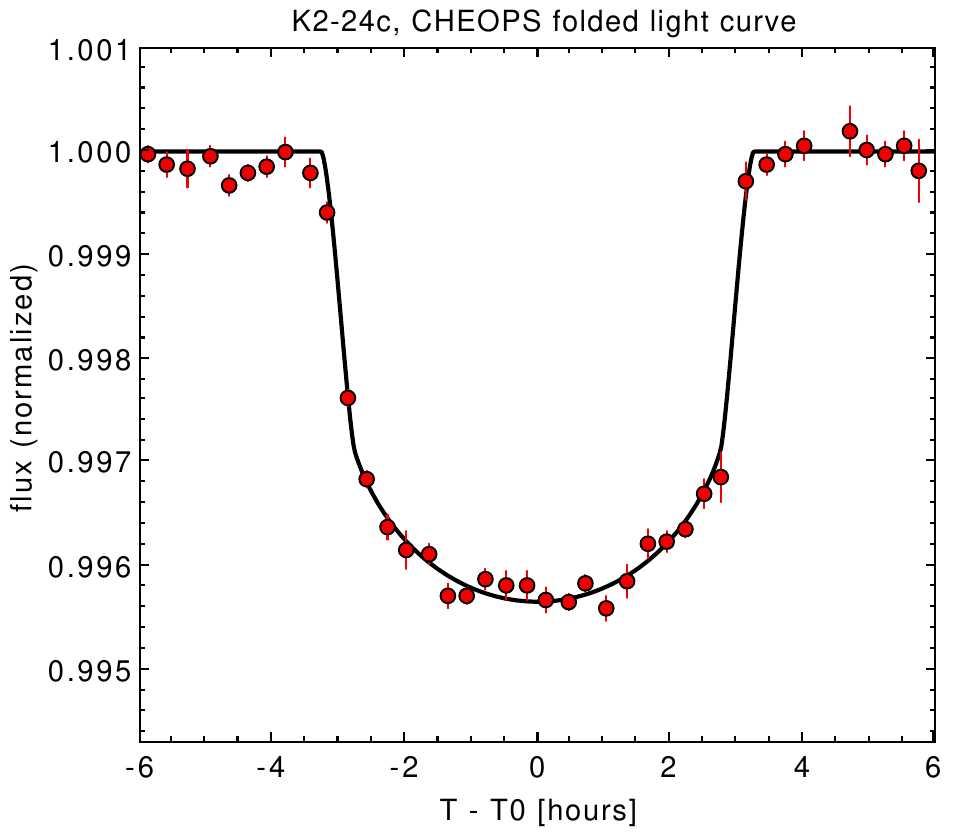}\\
    \caption{Folded light curves of K2-24b (left panel) and K2-24c (right panel) from CHEOPS, binned over 0.3-hour intervals.}
    \label{fig:lcbin}
\end{figure*}


It is evident that, particularly for planet c, some transits are partial ones. This is due to the transit predictions by \citetalias{Petigura2018} becoming increasingly inaccurate as the time passed since the K2/Spitzer observations increased. The $O-C$ (observed minus calculated) discrepancy is much larger than the prediction errors reported in Table~4 of \citetalias{Petigura2018}, demonstrating that their dynamical solution had to be revised and improved in order to reliably predict the future transit times.

\subsection{Radial velocities}\label{sec:observations_rv}

The merged data set we collected is made of 89 RV observations in total from three different instruments, including 63 RV points from HIRES, published by \citetalias{Petigura2018} (the first 32 having already been analyzed by \citetalias{Petigura2016}); 16 RV points from PFS, presented by \citet{Dai2016}; ten RV points from HARPS (PID: 095.C-0718), also presented by \citet{Dai2016}. Two additional HARPS points can be found in the ESO archive (PID: 191.C-0873), but we did not include them in our analysis since they are affected by an RV offset introduced on 2015-06-03.
We emphasize that our goal in this work was to fit all of these RV data simultaneously for the first time, as \citetalias{Petigura2018} did not include the PFS and HARPS data in their modeling.

\section{Light curve modeling}\label{sec:photometric}

We performed global modeling of our 13 CHEOPS, six K2, and one HST light curves by simultaneously fitting the signal of planet -b and -c on both data sets, using the \pyorbit{}\footnote{\url{https://github.com/LucaMalavolta/PyORBIT}} software version 10 \citep{Malavolta2016,Malavolta2018}. After some preliminary tests, we decided not to incorporate the Spitzer light curves into our global fit, as their transit depths are clearly inconsistent with the K2, CHEOPS, and HST data, and also with each other. This could be due to an imperfect correction of systematic errors since the transits are partial and the detrending process works by extrapolation rather than interpolation. To avoid any bias in our retrieved planetary parameters, we took the transit times $T_0$ for our dynamical analysis (Section~\ref{sec:dynamical}) from Table~1 of \citetalias{Petigura2018} instead.

The K2, CHEOPS, and HST transits were modeled with the Batman code \citep{Kreidberg2015} and parametrized as a function of the impact parameter $b$, the radius ratio $R_p/R_\star$, and the scaled semi-major axis $a/R_\star$. The transit model for K2 was super-sampled by a factor of ten to account for the non-negligible length of the K2 exposure times (30-minute cadence). Each transit time $T_0$ (13 from CHEOPS, six from K2, one from HST) was treated as a free, independent parameter, so the orbital period $P$ was fixed at its average value interpolated over our observations, that is, $P_b=20.8891$ and $P_c=42.3391$~days. The limb darkening effect was modeled through a quadratic law with two parameters called $u_1$ and $u_2$ for each instrument. Internally, $u_1$ and $u_2$ were re-parametrized as $q_1$ and $q_2$ following the prescription by \citet{Kipping2013} to minimize the correlation between the two parameters. 

The residual systematic errors present in the CHEOPS light curves $f(t)$ were detrended as a linear combination of terms as a function of external parameters: the first and second-order derivative of the centroid offset in $x$ and $y$ pixel coordinates ($\mathrm{d}f/\mathrm{d}x$, $\mathrm{d}^2f/\mathrm{d}x^2$, $\mathrm{d}f/\mathrm{d}y$,  $\mathrm{d}^2f/\mathrm{d}y^2$), background level ($\mathrm{d}f/\mathrm{d}b$), photometric contamination factor  ($\mathrm{d}f/\mathrm{d}\textrm{contam}$), the first three harmonics of the spacecraft roll angle (in $\cos\phi$ and $\sin\phi$), and a quadratic baseline  $f_0 + \mathrm{d}f/\mathrm{d}t + \mathrm{d}^2f/\mathrm{d}t^2$. The roll angle term, as expected with data from a low Earth orbit satellite, always dominates. Since adding 15 free parameters for each CHEOPS data set would have implied $13\times 15 = 195$ free parameters in our global fit just for the detrending (which is prohibitively expensive in terms of computational time), we went through the two-stage approach described in \citet{Nascimbeni2023}. In a first pass, each individual CHEOPS light curve was fitted with both a transit model and the detrending model. Then the detrended light curves were fed into the final global fitting. The latter therefore had 52 free parameters: six LD coefficients ($u_1$ and $u_2$ for CHEOPS, K2, HST), six planetary parameters ($b$, $R_p/R_\star$, $a/R_\star$ for -b and -c), 20 transit times, and 20 jitter parameters (one for each light curve). 

We set uninformative priors on all our fitting parameters. The only exceptions are the six limb darkening parameters $u_1$, $u_2$ for CHEOPS, K2, and HST. We carried out two independent analyses: the first one with fully uninformative priors (LD-free) and a second one by centering the prior at the theoretical value computed by the \texttt{LDTk} code \citep{Parviainen2015} and increasing to 0.05 the associated Gaussian error in order to accommodate for the well-known underestimation by models  (LD-prior). The input stellar parameters were computed by the CHEOPS Target Characterization (TS3) working group according to the procedure described by \citet{Borsato2021}, Section~3.2.1, and reported in Table~\ref{tab:stellar_param}. We set \pyorbit{} to use \texttt{PyDE} \citep{Parviainen2016} to find a reasonable starting point in the parameter space ($50\,000$ generations with a population size of $8\times N_\mathrm{par}$, where $N_\mathrm{par}$ is the number of free parameters). Then we initialized an Markov chain Monte Carlo (MCMC) optimization with \texttt{emcee} \citep{Foreman2013}, running for $500\,000$ steps and setting a thinning factor of 100. After discarding the first $50\,000$ steps as the burn-in phase, convergence was checked by auto-correlation function analysis (ACF). 

\begin{table}
    \centering\centering\renewcommand{\arraystretch}{1.2}
        \caption{Stellar parameters of K2-24 adopted for our analysis.}
    \begin{tabular}{ccc} \hline \hline
    Parameter [unit] & value & source \\ \hline
    distance [pc] & $170.5\pm 0.5$\phantom{00} & \emph{Gaia} DR3 \\
    $T_\mathrm{eff}$ [K] & $5726\pm 65$\phantom{00} & TS3 \\
    $\log g$ [cgs] & $4.48\pm 0.03$ & TS3 \\
    $[\mathrm{Fe}/\mathrm{H}]$ & $0.41\pm0.05$ & TS3 \\
    age~[Gyr]  & $4.9\pm1.7$ & TS3 \\
    $v\sin i$~[km/s] & <2 & TS3 \\
    $v_\mathrm{mic}$~[km/s]& $1.09\pm0.03$ & TS3 \\
    $R_\star$ [$R_\odot$] & $1.185\pm 0.011$ & TS3 \\
    $M_\star$ [$M_\odot$] & $1.091\pm 0.056$ & TS3 \\ \hline
    \end{tabular}\tablefoot{Sources: \emph{Gaia} DR3 \citep{GaiaDR3}, CHEOPS working group TS3 (see \citealt{Borsato2021}, Section 3.2.1 for details)}
    \label{tab:stellar_param}
\end{table}

All the final best-fit parameters of interest from the MCMC distributions are reported in Table~\ref{tab:pyorbit} for both the ``LD-free'' and ``LD-prior'' case; transit times are reported separately in Table~\ref{tab:t0}. The corresponding corner plots for the transit shape parameters (i.e., excluding transit times, LD, and jitter parameters) are shown in Fig.~\ref{fig:corner_b} and Fig.~\ref{fig:corner_c}. The best-fit values of $u_1$ and $u_2$ are consistent between the LD-free and LD-prior case (Table~\ref{tab:pyorbit}), although of course LD-prior has smaller error bars; all the remaining parameters agree within 1~$\sigma$. We adopt the LD-prior solution throughout the following analysis. We plot the CHEOPS light curves of K2-24b and -c folded on the best-fit individual $T_0$ and binned over 0.3-hour intervals in Fig.~\ref{fig:lcbin}.

\begin{table*}\centering\small\renewcommand{\arraystretch}{1.3}
\caption{Posterior and derived parameters of the K2-24 system from the global \pyorbit{} fit on the K2+HST+CHEOPS data set.}
\begin{tabular}{l|cc|cc|c}
\multicolumn{1}{c}{ } & \multicolumn{2}{c}{\emph{LD-prior}$^\dagger$ (adopted)} & \multicolumn{2}{c}{\emph{LD-free}$^\dagger$} & Literature \\
\hline\hline
 Parameter [unit] & MAP (HDI$\pm 1\sigma$) & Prior & MAP (HDI$\pm 1\sigma$) & Prior & \citetalias{Petigura2016} \\ \hline 
 \emph{Planet K2-24b} \rule{0pt}{12pt} & & & & & \\
 $a/R_\star$ & $30.38_{-1.00}^{+0.53}$  & $\mathcal{U}(20,40)$ & $30.18_{-1.30}^{+0.73}$  & $\mathcal{U}(20,40)$ &  $28.6_{-3.6}^{+1.7} $\\
 $R_p/R_\star$ & $0.04356_{-0.00021}^{+0.00034}$  & $\mathcal{U}(0.0,0.1)$ & $0.04368_{-0.00030}^{+0.00049}$  & $\mathcal{U}(0.0,0.1)$& $0.0431_{-0.0008}^{+0.0017}$ \\
 $b$ & $0.20_{-0.13}^{+0.12}$  & $\mathcal{U}(0,1)$ & $0.24_{-0.15}^{+0.13}$  & $\mathcal{U}(0,1)$ & $0.37_{-0.24}^{+0.22}$ \\
 $i$ [deg] & $89.63_{-0.25}^{+0.25}$  & (derived) & $89.55_{-0.28}^{+0.30}$  & (derived) & $89.25_{-0.61}^{+0.49}$ \\
 $R_p/R_\oplus$ & $5.638_{-0.061}^{+0.065}$  & (derived) & $5.655_{-0.069}^{+0.076}$  & (derived) & $5.68\pm 0.56$\\
 $T_{14}$ [days] & $0.22437_{-0.00075}^{+0.00086}$  & (derived) & $0.22419_{-0.00083}^{+0.00099}$  & (derived) & $0.2283_{-0.0017}^{+0.0029}$ \\
 $T_{23}$ [days] & $0.20470_{-0.00094}^{+0.00078}$  & (derived) & $0.2042_{-0.0013}^{+0.0010}$  & (derived) & $0.2062_{-0.0046}^{+0.0020}$ \\
 $a$ [au] & $0.1673_{-0.0056}^{+0.0034}$  & (derived) & $0.1662_{-0.0071}^{+0.0043}$  & (derived) & $0.154\pm 0.002$ \\ \hline
 \emph{Planet K2-24c} \rule{0pt}{16pt} & & & & & \\
 $a/R_\star$ & $47.6_{-2.0}^{+2.4}$  & $\mathcal{U}(30,60)$ & $46.9_{-1.7}^{+2.1}$  & $\mathcal{U}(30,60)$ & $52.6_{-2.6}^{+0.0}$ \\
 $R_p/R_\star$ & $0.06133_{-0.00080}^{+0.00068}$  & $\mathcal{U}(0.0,0.1)$ & $0.06171_{-0.00080}^{+0.00068}$  & $\mathcal{U}(0.0,0.1)$ & $0.0594_{-0.0004}^{+0.0010}$\\
 $b$ & $0.462_{-0.100}^{+0.066}$  & $\mathcal{U}(0,1)$ & $0.489_{-0.083}^{+0.056}$  & $\mathcal{U}(0,1)$ & $0.22_{-0.16}^{+0.17}$\\
 $i$ [deg] & $89.44_{-0.11}^{+0.15}$  & (derived) & $89.40_{-0.10}^{+0.12}$  & (derived) & $89.76_{-0.21}^{+0.18}$\\
 $R_p/R_\oplus$ & $7.93_{-0.13}^{+0.12}$  & (derived) & $7.98_{-0.13}^{+0.12}$  & (derived) & $7.82\pm 0.72$\\
 $T_{14}$ [days] & $0.2707_{-0.0018}^{+0.0018}$  & (derived) & $0.2708_{-0.0016}^{+0.0016}$  & (derived) & $0.2696_{-0.0013}^{+0.0017}$ \\
 $T_{23}$ [days] & $0.2315_{-0.0025}^{+0.0026}$   & (derived) & $0.2301_{-0.0025}^{+0.0028}$   & (derived) & $0.2375_{-0.0025}^{+0.0013}$ \\
 $a$ [au] & $0.262_{-0.011}^{+0.014}$  & (derived) & $0.258_{-0.010}^{+0.012}$  & (derived) & $0.247\pm0.004$ \\ \hline
 \emph{Limb darkening parameters} \rule{0pt}{16pt} & & & & & (fixed:) \\
 $u_1$ (K2, linear) & $0.527_{-0.025}^{+0.025}$  & $\mathcal{N}(0.55,0.05)$  & $\phantom{-}0.561_{-0.063}^{+0.065}$  & $\mathcal{U}(0.0,1.0)$ & $0.568\pm 0.003$ \\
 $u_2$ (K2, quadratic) & $0.043_{-0.039}^{+0.040}$  & $\mathcal{N}(0.05,0.05)$ & $-0.02_{-0.10\phantom{0}}^{+0.11}\phantom{0}$  & $\mathcal{U}(-0.5,0.5)$ & $0.098\pm 0.005$ \\
$u_1$ (HST, linear) & $0.196_{-0.036}^{+0.035}$  & $\mathcal{N}(0.24,0.05)$  & $\phantom{-}0.103_{-0.070}^{+0.098}$  & $\mathcal{U}(0.0,1.0)$ & $0.568\pm 0.003$ \\
 $u_2$ (HST, quadratic) & $0.147_{-0.046}^{+0.046}$  & $\mathcal{N}(0.16,0.05)$ & $\phantom{-}0.24_{-0.20}^{+0.16\phantom{0}}\phantom{0}$  & $\mathcal{U}(-0.5,0.5)$ & $0.098\pm 0.005$ \\
 $u_1$ (CHEOPS, linear) & $0.562_{-0.030}^{+0.030}$  & $\mathcal{N}(0.55,0.05)$ & $\phantom{-}0.652_{-0.097}^{+0.092}$  & $\mathcal{U}(0.0,1.0)$ & --- \\
 $u_2$ (CHEOPS, quadratic) & $0.035_{-0.041}^{+0.041}$  & $\mathcal{N}(0.06,0.05)$  & $-0.11_{-0.13}^{+0.15\phantom{0}}\phantom{0}$  & $\mathcal{U}(-0.5,0.5)$  & ---\\ \hline
\end{tabular}\tablefoot{The columns give the parameter name and unit (where applicable); the MAP value of the posterior distribution; its 1-$\sigma$ HDI and the adopted prior for the LD-prior and LD-free case, respectively ($\dagger$ see Section~\ref{sec:photometric} for details); and the best-fit values from the literature \citepalias{Petigura2016}, for comparison.}
\label{tab:pyorbit}
\end{table*}

\begin{table}
    \centering\centering\small\renewcommand{\arraystretch}{1.2}
        \caption{Transit times of K2-24b and K2-24c measured or adopted for our dynamical analysis.}
    \begin{tabular}{cccc} \hline \hline
    Planet & Instrument & $T_0$ [$\textrm{BJD}_\textrm{TDB}$] & Source \\ \hline
  -b & K2      &   $2456905.79529 \pm 0.00035$ &     this work \\
  -c & K2      &   $2456915.62477 \pm 0.00021$ &     this work \\
  -b & K2      &   $2456926.67909 \pm 0.00049$ &     this work \\
  -b & K2      &   $2456947.56565 \pm 0.00035$ &     this work \\
  -c & K2      &   $2456957.98825 \pm 0.00019$ &     this work \\
  -b & K2      &   $2456968.45070 \pm 0.00033$ &     this work \\
  -b & Spitzer &   $2457323.61610 \pm 0.00110$ &      \citetalias{Petigura2018}      \\
  -c & Spitzer &   $2457339.00020 \pm 0.00140$ &      \citetalias{Petigura2018}      \\
  -c & Spitzer &   $2457550.50740 \pm 0.00150$ &      \citetalias{Petigura2018}      \\
  -b & Spitzer &   $2457553.50490 \pm 0.00160$ &      \citetalias{Petigura2018}      \\
  -b & HST     &   $2457574.40950 \pm 0.00350$ &     this work      \\
  -b & CHEOPS  &   $2459391.82076 \pm 0.00064$ &     this work, \texttt{b1} \\
  -b & CHEOPS  &   $2459412.70870 \pm 0.00125$ &     this work, \texttt{b2} \\
  -c & CHEOPS  &   $2459413.46115 \pm 0.00094$ &     this work, \texttt{c1} \\
  -c & CHEOPS  &   $2459667.64590 \pm 0.00175$ &     this work, \texttt{c2} \\
  -b & CHEOPS  &   $2459705.07047 \pm 0.00082$ &     this work, \texttt{b3} \\
  -c & CHEOPS  &   $2459710.02459 \pm 0.00054$ &     this work, \texttt{c3} \\
  -b & CHEOPS  &   $2459725.94986 \pm 0.00062$ &     this work, \texttt{b4} \\
  -b & CHEOPS  &   $2459746.83330 \pm 0.00320$ &     this work, \texttt{b5} \\
  -c & CHEOPS  &   $2459752.39985 \pm 0.00093$ &     this work, \texttt{c4} \\
  -c & CHEOPS  &   $2460049.00500 \pm 0.00076$ &     this work, \texttt{c5} \\
  -c & CHEOPS  &   $2460091.36711 \pm 0.00105$ &     this work, \texttt{c6} \\
  -b & CHEOPS  &   $2460101.84061 \pm 0.00066$ &     this work, \texttt{b6} \\
  -b & CHEOPS  &   $2460122.73033 \pm 0.00066$ &     this work, \texttt{b7} \\ \hline
    \end{tabular}\tablefoot{The columns give the following: planet name (K2-24b or K2-24c), instrument used, best-fit transit time $T_0$ in the BJD-TDB standard \citep{Eastman2010} along with its 1-$\sigma$ error bar (symmetrized), and the source (plus the light curve ID from Table~\ref{tab:log} for the CHEOPS data). All the $T_0$ values from this work were obtained through the global photometric fit described in Section~\ref{sec:photometric}.}
    \label{tab:t0}
\end{table}

In the last column of Table~\ref{tab:pyorbit} we also compare our results with the literature, namely with \citetalias{Petigura2016} (\citetalias{Petigura2018} did not present a new set of independent planetary parameters since the work was based on priors from \citetalias{Petigura2016}). Overall there is very good agreement. The planetary radii $R_b$ and $R_c$, in particular, are consistent within 1~$\sigma$, but our error bars are improved by an order of magnitude (i.e., from a relative error of $\sim9$\% to $\sim1$\%). The uncertainty is now limited by our current knowledge of the stellar radius $\sigma(R_\star)/R_\star\simeq 1\%$ (Table~\ref{tab:stellar_param}).

\section{Dynamical modeling}\label{sec:dynamical}

We carried out a dynamical modeling of the K2-24 system and its strong TTV signals by simultaneously fitting the three RV data sets available (see Section~\ref{sec:observations_rv}) and the transit times ($T_{0}$s) extracted with \pyorbit{} (see Table~\ref{tab:t0}) through the \trades{} code\footnote{\url{https://github.com/lucaborsato/trades}}
\citep{Borsato2014, Borsato2019, Borsato2021}.
We adopted a parameterization similar to \citet{Nascimbeni2023}, assuming a three-planet\footnote{Candidate planet d (at $\simeq 1.1$~au) is far enough to be dynamically decoupled from the inner pair, so in principle it should not impact the TTV signals. It has to be included in our dynamical modeling, though, because of its effect on RV,
as in \citetalias{Petigura2018}.} model and fitting for the stellar mass $M_\star$, planetary-to-star mass ratio $M_\mathrm{p}/M_\star$, periods $P$, mean longitude\footnote{The mean longitude is defined as $\lambda=\mathcal{M}+\Omega+\omega$, where $\mathcal{M}$ is the mean anomaly, $\Omega$ is the longitude of ascending node, and $\omega$ is the argument of pericenter.} $\lambda$ of all planets, eccentricity $e$, and argument of periastron passage $\omega$ in the form $\sqrt{e}\cos\omega$ and $\sqrt{e}\sin\omega$ for planets -b and -c (as specified by the indexes $b$ and $c$). We also fit a jitter term in $\log_{2}$-space, and an offset for each RV data set.
We fixed the following parameters: longitude of ascending node $\Omega$ to $180\degr$ for each planet; circular orbit of planet d (eccentricity $e_\mathrm{d}=0$ and argument of periastron $\omega_\mathrm{d}=90\degr$); and inclination, $i$, of planets -b and -c as in Table~\ref{tab:pyorbit} and to $90\degr$ for planet d. All the parameters are defined at the reference time 
$T_\mathrm{ref} = 2\,456\,905\ \mathrm{BJD_{TDB}}$. We defined the parameter priors in the physical space and converted them into fitting space; all the priors used are reported in Table~\ref{tab:trades}.

\begin{figure*}
    \centering
    \includegraphics[width=\columnwidth]{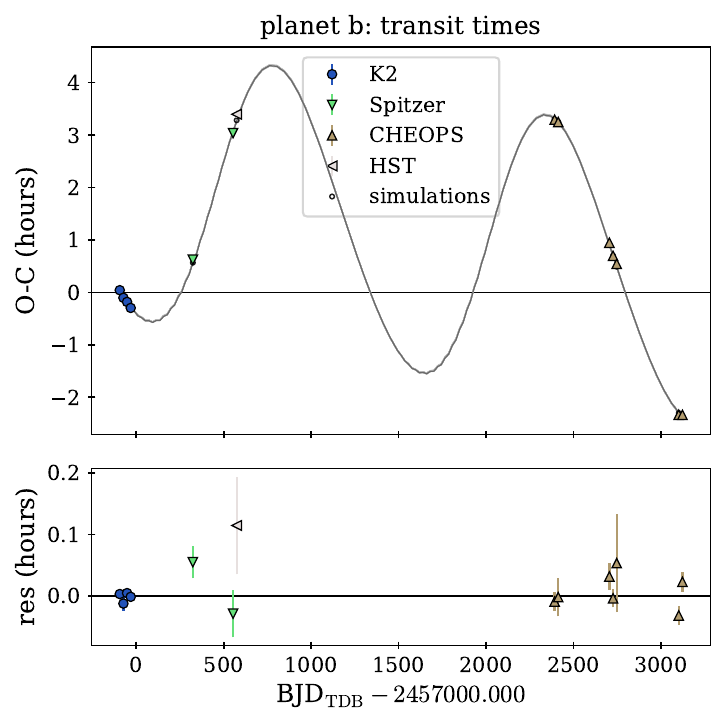} \includegraphics[width=\columnwidth]{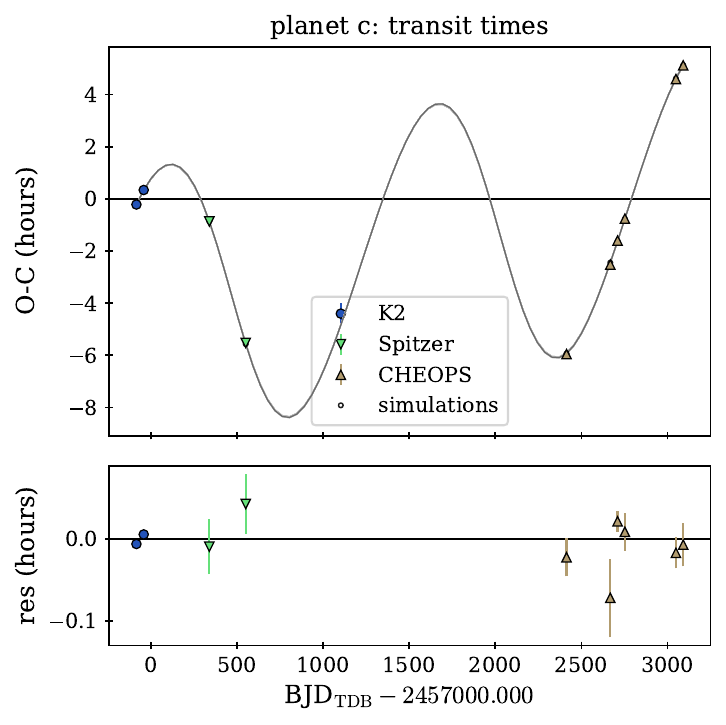}
    \caption{TTV modeling of K2-24b (left panels) and K2-24c (right panels). Top panel: $(O-C)$ diagram, calculated by subtracting the $T_0$ predicted by the linear ephemeris to the observed transit times (K2, Spitzer, HST, CHEOPS). The $O-C$ values computed from the observed $T_0$s are plotted 
    \valerio{with different solid colors and marker shapes}, while the $O-C$ values computed at the same epochs by the best-fit \trades{} dynamical model (Section~\ref{sec:dynamical}) are plotted as \valerio{black open} circles. Samples drawn from the posterior distribution from \trades{} within HDI are shown as gray lines. 
    Bottom panel: Residuals computed as the difference between observed and simulated $T_0$s.}
    \label{fig:oc_b}
\end{figure*}

\begin{figure*}
    \sidecaption
    \includegraphics[width=1.15\columnwidth, trim=0 11 0 0]{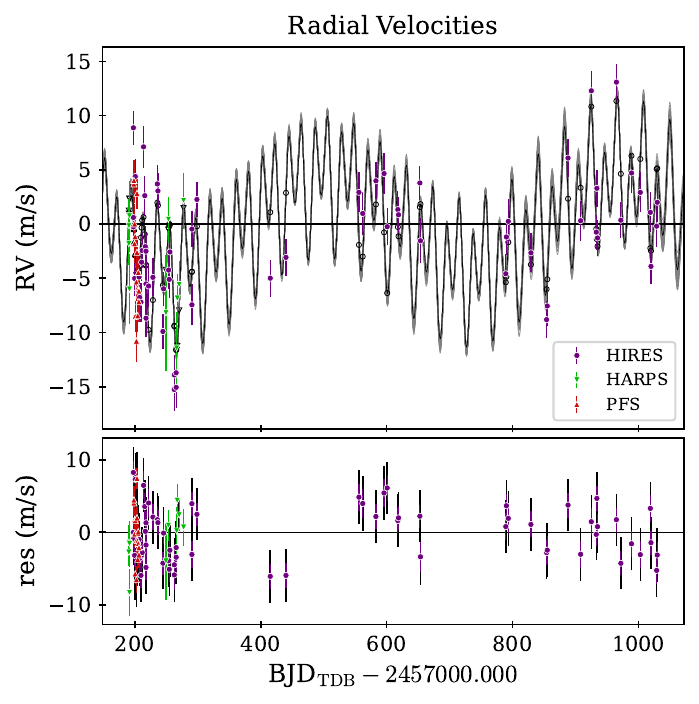}
    \caption{Radial velocities of K2-24. Upper panel: RV plot; each data set (HIRES, HARPS, PFS) is plotted with a different marker shape and color, as in legend. 
    The MAP RV model from the best-fit TRADES orbital solution is plotted as a black line, while the RV points computed at the observed epochs are rendered with open black circles.
    Samples drawn from the posterior distribution within HDI are shown as gray lines.
    Lower panel: RV residuals with respect to the TRADES RV best-fit model. 
    The corresponding jitter determined from the best-fit model has been added in quadrature to the measured uncertainty of each data point.}
    \label{fig:rv}
\end{figure*}

We first ran \trades{} with \pyde{} (100 different configurations for $150\, 000$ generations) to find a suitable starting point. Then, we ran \emcee{} with 100 walkers for $1\, 000\, 000$ steps, and we applied a conservative thinning factor of 100.
As in \citet{Nascimbeni2023},  we used a combination of the differential evolution proposal \citep[80\% of the walkers;][]{DEMOVE2014ApJS..210...11N} and the snooker differential evolution proposal \citep[20\% of the walkers;][]{terBraak2008}
as the sampler within \emcee{}. After checking the chains' convergence through Gelman-Rubin statistics \citep{GelmanRubin1992}, 
Geweke criterion \citep{geweke1991}, ACF, and visual inspections, we discarded as burn-in the first $50\%$ of the steps.
From the posterior distributions, we extracted the maximum a posteriori (MAP\footnote{By MAP we mean the set of parameters that maximize the log-probability of the posterior distributions. If all the priors were uninformative and uniform, then the MAP would correspond to the maximum likelihood estimation (MLE).})
as the best-fit parameters and the uncertainties as the high density interval (HDI) at $68.27\%$.\footnote{An 
HDI at $68.27\%$ is equivalent to the $16^\mathrm{th}$--$84^\mathrm{th}$ percentiles of a Gaussian distribution.}
The best-fit parameters from \trades{} and their uncertainties are reported in Table~\ref{tab:trades},
with a comparison with \citetalias{Petigura2018} for the parameters in common.
The TTV and RV models from the best-fit orbital solution by \trades{} were also plotted along with
the observed data points in Fig.~\ref{fig:oc_b} and Fig.~\ref{fig:rv}, respectively. 
The fit looks perfectly satisfactory, with an overall reduced $\chi^2$ (TTV+RV) of 1.33 with 94 degrees of freedom. The corresponding $\ln\mathcal{L}$ and $\ln(\textrm{probability})$ values are $-107.242$ and $-107.623$, respectively.
\par

\label{sec:predictions}

A useful application of our dynamical model is the prediction of future transit events for any follow-up opportunity. 
As for planets -b and  c, transits cannot be reliably scheduled according to a linear ephemeris. The combined impact of a poorly constrained linear ephemeris with a set of orbital parameters determined over a relatively short time span has been discussed in \citet{Borsato2022} in the context of the scientific preparation of the Ariel mission \citep{Tinetti2018}.

All the transits of planets -b and -c predicted by our best-fit dynamical model up to and including the year 2029
are reported along with the associated uncertainty in 
electronic form at CDS.
The predicted transit times and their associated uncertainties were calculated by integrating 100 orbital solutions randomly chosen from the \trades{} fit posterior and then computing the median and the 68.27\% HDI interval at each transit epoch.

\begin{table*}\centering\small\renewcommand{\arraystretch}{1.25}
\caption{Posterior and derived parameters of the K2-24 system from the dynamical fit with TRADES (Section~\ref{sec:dynamical}).}
\begin{tabular}{l|cc|c}
\hline\hline
 Parameter [unit] & MAP (HDI$\pm 1\sigma$) & Prior & \citetalias{Petigura2018} \\ \hline 
 \emph{Host star K2-24} \rule{0pt}{16pt} & & & \\
 $M_\star\, [M_\odot]$ & $1.041_{-0.016}^{+0.067}$ & $\mathcal{G}(1.091, 0.056)$ & $1.07\pm 0.06$* \\ 
 \hline
 \emph{Planet K2-24b} \rule{0pt}{16pt} & & & \\
 $M_\mathrm{b}/M_\star$ & $\left (59_{-1}^{+2} \right )\cdot 10^{-6}$  & $\mathcal{U}(8.8\cdot 10^{-8},2.8\cdot 10^{-3})$ &  --- \\ 
 $M_\mathrm{b}/M_\oplus$ & $20.6_{-0.3}^{+1.6}$  & (derived) &  $19.0_{-2.1}^{+2.2}$ \\ 
 $P_b$ at $T_\mathrm{ref}$ [days]  & $20.88257_{-0.00007}^{+0.00005}$  & $\mathcal{U}(18,22)$ &  --- \\ 
$\left\langle P_b \right\rangle^\mathsection$ [days] & $20.8895961 \pm 0.0000096^\mathsection$  & (derived) &  $20.88977_{-0.00035}^{+0.00034}$ \\ 
 $\sqrt{e_b}\cos\omega_b$ & $0.221_{-0.004}^{+0.003}$  & $\mathcal{U}(-\!\sqrt{0.5},\sqrt{0.5})$ &  --- \\ 
 $\sqrt{e_b}\sin\omega_b$ & $-0.0318_{-0.0016}^{+0.0013}$  & $\mathcal{U}(-\!\sqrt{0.5},\sqrt{0.5})$ &  --- \\ 
 $\lambda_b$ [deg] & $250.61_{-0.13}^{+0.22}$  & $\mathcal{U}(0,360)$ &  --- \\ 
  $\mathcal{M}_b$ [deg]  & $78.8_{-0.5}^{+0.8}$  & (derived) &  --- \\ 
 $e_b$  & $0.0498_{-0.0018}^{+0.0011}$  & (derived) &  $0.06\pm 0.01$ \\ 
 $\omega_b$ & $351.8_{-0.5}^{+0.4}$  & (derived) &  --- \\ 
 $\rho_b$ [cgs] & $0.632_{-0.023}^{+0.054}$  & (derived) &  $0.64_{-0.10}^{+0.12}$ \\ 
 $T_\mathrm{eq}$ [K] & $732\pm 11$  & (derived) &  --- \\
 TSM & $62\pm 5$  & (derived) &  --- \\
 \hline
 \emph{Planet K2-24c} \rule{0pt}{16pt} & & & \\
 $M_\mathrm{c}/M_\star$ & $\left (47.3_{-0.8}^{+1.8} \right )\cdot 10^{-6}$  & $\mathcal{U}(8.8\cdot 10^{-8},2.8\cdot 10^{-3})$ &  --- \\ 
 $M_\mathrm{c}/M_\oplus$ & $16.4_{-0.2}^{+1.3}$  & (derived) &  $15.4_{-1.8}^{+1.9}$ \\ 
 $P_c$ at $T_\mathrm{ref}$ [days] & $42.3773_{-0.0004}^{+0.0005}$  & $\mathcal{U}(40,44)$ &  --- \\ 
 $\left\langle P_c \right\rangle^\mathsection$ [days] & $42.34732 \pm 0.00018^\mathsection$  & (derived) &  $42.3391\pm0.0012$ \\ 
 $\sqrt{e_c}\cos\omega_c$ & $-0.1268_{-0.0012}^{+0.0058}$  & $\mathcal{U}(-\!\sqrt{0.5},\sqrt{0.5})$ &  --- \\ 
 $\sqrt{e_c}\sin\omega_c$ & $0.110_{-0.002}^{+0.004}$  & $\mathcal{U}(-\!\sqrt{0.5},\sqrt{0.5})$ &  --- \\ 
 $\lambda_c$ [deg] & $182.15_{-0.13}^{+0.03}$  & $\mathcal{U}(0,360)$ &  --- \\ 
 $\mathcal{M}_c$ [deg]  & $223.1_{-0.6}^{+2.3}$  & (derived) &  --- \\ 
 $e_c$  & $0.0282_{-0.0007}^{+0.0003}$  & (derived) & $<0.07$ (90\% conf.)$^\dagger$ \\ 
 $\omega_c$ & $139.0_{-2.5}^{+0.6}$  & (derived) &  --- \\  
 $\rho_c$ [cgs] & $0.181_{-0.009}^{+0.017}$  & (derived) &  $0.20_{-0.03}^{+0.04}$ \\ 
  $T_\mathrm{eq}$ [K] & $591\pm 12$  & (derived) &  --- \\
 TSM & $177\pm 16$  & (derived) &  --- \\
 \hline
\emph{Candidate K2-24d} \rule{0pt}{16pt} & & & \\
 $M_\mathrm{d}/M_\star$ & $\left (155_{-13}^{+19} \right )\cdot 10^{-6}$  & $\mathcal{U}(8.8\cdot 10^{-8},2.8\cdot 10^{-3})$ &  --- \\ 
 $M_\mathrm{d}/M_\oplus$ & $54_{-4}^{+9}$  & (derived) &  $52\pm 14$ \\ 
 $P_d$ [days] & $469_{-15}^{+10}$  & $\mathcal{U}(100,1000)$ &  440?$^\ddagger$ \\ 
 $\lambda_d$ [deg] & $83_{-12}^{+15}$  & $\mathcal{U}(0,360)$ &  --- \\ 
 $\mathcal{M}_d$ [deg]  & $173_{-12}^{+15}$  & (derived) &  --- \\ 
 $e_d$  & 0.00  & (circular) &  (circular) \\ \hline
 \emph{Jitter and offset terms} \rule{0pt}{16pt} & & & \\
$\gamma_\mathrm{HIRES}$ [m/s] & $1.1_{-0.4}^{+0.4}$ &  $\mathcal{U}(-5000,5000)$ & --- \\
$\gamma_\mathrm{HARPS}$ [m/s] & $763.1_{-0.9}^{+0.6}$ &  $\mathcal{U}(-5000,5000)$ & --- \\
$\gamma_\mathrm{PFS}$ [m/s] & $5.3_{-1.0}^{+0.8}$ &  $\mathcal{U}(-5000,5000)$ & --- \\
$j_\mathrm{HIRES}$ [m/s] & $3.2_{-0.2}^{+0.4}$ & $\mathcal{U}(10^{-15}, 100)$ & --- \rule{0pt}{16pt}\\
$j_\mathrm{HARPS}$ [m/s] & $0.0_{-0.0}^{+0.0}$ & $\mathcal{U}(10^{-15}, 100)$ & --- \\ 
$j_\mathrm{PFS}$ [m/s] & $3.3_{-0.6}^{+0.6}$ & $\mathcal{U}(10^{-15}, 100)$ & --- \\ \hline
\end{tabular}\tablefoot{All the parameters were computed at the reference time $T_\mathrm{ref} = 2\,456\,905\ \mathrm{BJD_{TDB}}$. The columns give the parameter name, the MAP value of the posterior distribution and its 1-$\sigma$ HDI, the adopted prior, and the best-fit values from \citetalias{Petigura2018} for comparison (when available). *Derived from spectroscopy; not a fitted parameter within the \citetalias{Petigura2018} dynamical model. $^\mathsection$Average orbital period corresponding to the MAP model computed by TRADES. $^\dagger$The value of $e_c$ from \citetalias{Petigura2018} was derived by assuming a prior based on the distribution of $\langle e\rangle$ observed
in \emph{Kepler} multi-planet systems. $^\ddagger$The best-fit period $P_d$ and its associated uncertainty are not explicitly reported by \citetalias{Petigura2018}.}\label{tab:trades}
\end{table*}

\section{Discussion and conclusions}\label{sec:conclusions}

In our work, we have merged all the available space-based photometry of K2-24b and -c (including 13 unpublished CHEOPS light curves; Section~\ref{sec:observations}) and derived improved stellar parameters for K2-24 (Table~\ref{tab:stellar_param}) to perform a global transit fit (Section~\ref{sec:photometric}), which yielded a homogeneous set of planetary parameters and transit times (Tables~\ref{tab:pyorbit}, \ref{tab:t0}). Then we fitted the latter together with all the available RVs (HIRES, PFS, HARPS) through an RV plus TTV dynamical model (Section~\ref{sec:dynamical}) in order to get a complete orbital solution for K2-24b and -c, and for candidate planet d as well (Table~\ref{tab:trades}). 

\subsection{Planetary parameters of K2-24b and K2-24c}\label{sec:dynamics}

All the derived parameters for planets -b and -c appear statistically consistent, at least within 2~$\sigma$, with those published by \citetalias{Petigura2018}, but they mostly have smaller error bars due to the increased S/N of the combined data set, the improved stellar parameters, and the much larger baseline of the observations. This is particularly true for the planetary radii ($R_b/R_\oplus=5.64\pm 0.06$, $R_c/R_\oplus=7.93\pm 0.12$) and masses ($M_b/M_\oplus=20.6_{-0.3}^{+1.6}$, $M_c/M_\oplus=16.4_{-0.2}^{+1.3}$), for which we reached a relative error of 1\% and 4-5\%, respectively. We confirm the unusually low density of the outer planet ($\rho_c = 0.181_{-0.009}^{+0.017}$~g~cm$^{-3}$), implying a very large gaseous envelope, possibly larger than 50\% and hence challenging a core-accretion scenario due to the onset of runaway accretion \citepalias{Petigura2018}.
Several alternative scenarios have been proposed to explain the existence of such ``super-puff'' planets \citep{Gao2020}, including light scattering from high-altitude photo-chemical hazes \citep{Ohno2021} or the presence of planetary rings on specific configurations \citep{Piro2020}. These hypotheses, however, require follow-up by JWST to be tested.

A particularly interesting variable to discuss is the orbital eccentricity, due to its important consequences on the planetary migration mechanisms. We measured an extremely significant non-zero eccentricity for both planets $e_b=0.049_{-0.002}^{+0.001}$, $e_c=0.0282_{-0.0007}^{+0.0003}$, confirming the findings by \citetalias{Petigura2018}, who based on the \citet{Lithwick2012} theory predicted that $e_b$ and $e_c$ cannot both be zero. It is worth noting that our best-fit values are perfectly compatible with their constraints, even though our analysis is based on uninformative priors only and does not adopt any assumption on the distribution of the eccentricity in the \emph{Kepler} population. We also mention that compared with the prediction of \citet{Antoniadou2020}, we found the eccentricity of -c to be at the very limit they set ($e_c< 0.05$). 

\subsection{\valerio{Dynamical stability}}
\label{sec:stability}

The K2-24 system hosts three planets (Table~\ref{tab:trades}), including two Neptune-mass planets ($M_b \approx 20.6$~$M_\oplus$, $M_c \approx 16.4$~$M_\oplus$) in the vicinity of a 2:1 mean motion resonance ($P_c / P_b \approx 2.029$).
Following a suggestion from the referee, we first checked the dynamical stability of our orbital solution by computing the angular momentum deficit \citep[AMD;][]{Laskar1997A&A...317L..75L, Laskar2000PhRvL..84.3240L, LaskarPetit2017A&A...605A..72L}  of the whole posterior distribution. Then we explored the stability by evaluating the AMD-Hill criterion proposed in Eq.~26 of \citet{Petit2018A&A...617A..93P}. We found that the whole posterior is AMD-Hill stable. We also ran an N-body integration with the Mean Exponential Growth factor of Nearby Orbits \citep[MEGNO;][]{Cincotta2000A&AS..147..205C} indicator through the \textsc{rebound} package with the \texttt{whfast} integrator \citep{rebound, reboundwhfast, wh, reboundvar}. We set a step size equal to $10\%$ of the shorter period of the system and integrated for $10^5$ years. We found that not only is the MAP solution stable ($\mathrm{MEGNO} = 2$) but that 1000 samples randomly selected from the posterior are also stable with $\mathrm{MEGNO} \simeq 2$.

\begin{figure*}
\sidecaption
    \centering
	\includegraphics[width=1.2\columnwidth, trim={30 15 0 0}, clip]{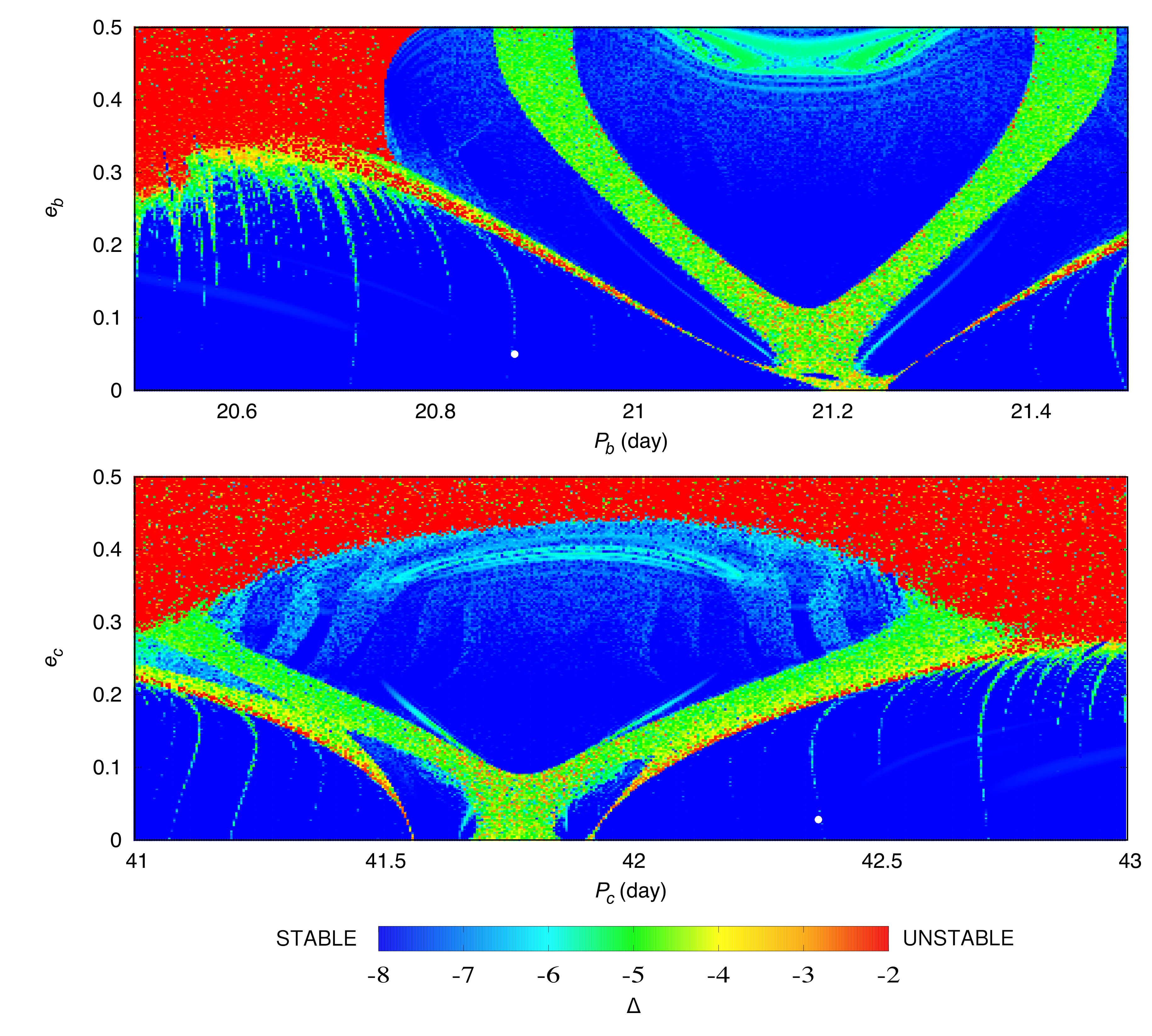}
    \caption{Stability analysis of the K2-24 planetary system. For fixed initial conditions (Table~\ref{tab:trades}), the parameter space of the system was explored by varying the orbital period and the eccentricity of planet -b (top panel) and of planet -c (bottom panel). The step size is $0.0025$ in the eccentricities, $0.0025$~d in the orbital period of planet $b$, and $0.005$~d in the orbital period of planet $c$. For each initial condition, the system was integrated over $10^4$~yr, and a stability indicator was calculated, which involved a frequency analysis of the mean longitude of the inner planet. The chaotic diffusion was measured by the variation in the frequency (see text). Red points correspond to highly unstable orbits, while blue points correspond to orbits that are likely to be stable on gigayear timescales. The white dots show the values of the best-fit solution (Table~\ref{tab:trades}).}
    \label{figSA}
\end{figure*}

\subsection{\valerio{Discussion of whether K2-24b and K2-24c are on an MMR configuration}}
\label{sec:mmr}

To get a wider view on the stability of the system \valerio{and to assess whether the system is truly on a resonant configuration,} we performed a dynamical analysis in a manner similar to other planetary systems \citep[e.~g.,][]{Correia_etal_2005, Correia_etal_2010}.
The system was integrated on a regular 2D mesh of initial conditions around the best fit, including planet d (Table~\ref{tab:trades}).
Each initial condition was integrated for $10^4$~yr using the symplectic integrator SABAC4 \citep{Laskar_Robutel_2001} with a step size of $10^{-3} $~yr and general relativity corrections.
Then, we performed a frequency analysis \citep{Laskar_1990, Laskar_1993PD} of the mean longitude of the inner planet over two consecutive time intervals of 5000~yr, and we determine the main frequency, $n$ and $n'$, respectively.
The stability was measured by $\Delta = |1-n'/n|$, which estimates the chaotic diffusion of the orbits.
In Fig.~\ref{figSA}, the results \valerio{for planet -b (top panel) and planet -c (bottom panel)} are reported in color: orange and red represent strongly chaotic and unstable trajectories; yellow indicates the transition between stable and unstable regimes; green corresponds to moderately chaotic trajectories but that are stable on gigayear timescales; and cyan and blue give extremely stable quasi-periodic orbits.
The best-fit solution obtained from our analysis (Table~\ref{tab:trades}) is marked with a white circle.

\begin{figure}
    \centering
	\includegraphics[width=\columnwidth]{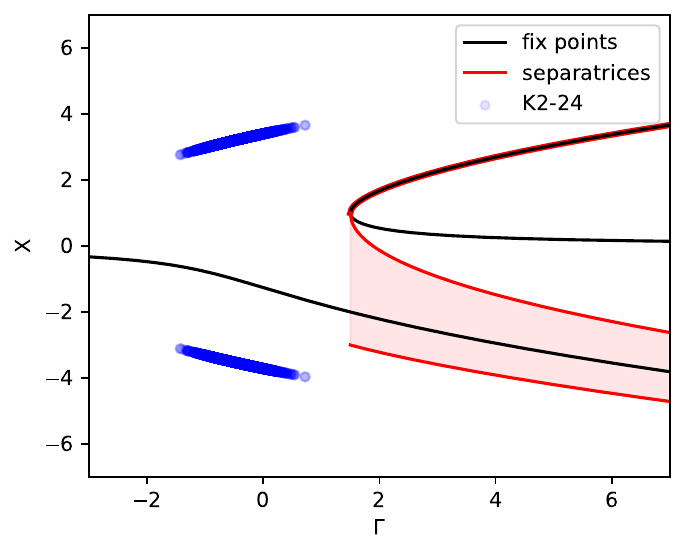}
    \caption{\valerio{One-degree of freedom model of the 2:1 MMR \citep{HenrardLemaitre1983,Deck2013}. We note that $X$ and $\Gamma$ are functions of the orbital elements and the masses of the system. The term $\Gamma$ represents how deep the system is in the resonance, while $X$ parameterizes the position of the fix points and separatrices. The blue dots represent the intersection of $10\,000$ randomly selected samples of the posterior with the $X$-$\Gamma$ plane.}}
    \label{fig:phasespace}
\end{figure}

We observed that the best-fit solution from Table~\ref{tab:trades} is completely stable, even if we increase the eccentricities up to 0.4. However, for eccentricities up to 0.1, which include the current best-fit determination ($e_b \approx 0.05$, $e_c \approx 0.03$), we observe that the system is outside the 2:1 mean motion resonance, which corresponds to the large stable structure above the V-shape chaotic region in the middle of the figures. 
\valerio{The TTV analysis constrained the resonant part of the architecture, as can be seen in Fig. \ref{fig:phasespace}. In that figure, we observed that the posterior of the fit lies outside the formal resonant domain (red area), unlike, for example, TOI-216, \emph{Kepler}-1705, or \emph{Kepler}-1972 \citep{Nesvorny2022,Leleu+2022}.}
We conclude that the K2-24 three-planet orbital solution presented in Table~\ref{tab:trades} is not in resonance but still reliable and supple to the uncertainties in the determination of the eccentricities of the two innermost planets.

We also note that the system is on the correct side of the resonance predicted by planetary migration models \citep[e.g.,][]{Lissauer_etal_2011K}.
This feature is usually attributed to tidal interactions with the parent star \citep[e.g.,][]{Delisle_Laskar_2014}, but in this case this mechanism does not seem to be very efficient because the orbital period of the inner Neptune-like planet is much longer than five days \citep{Correia_etal_2020}.

\valerio{Finally, with the best-fit solution, we monitored the evolution over $10\,000$ years of some parameters of the inner pair, including the period ratio $P_c/P_b$, the difference between the arguments of the pericenter $\Delta\omega$, and the critical resonant angles $\phi_1$, $\phi_2$  (Fig.~\ref{fig:evolution}). Interestingly, $\phi_1$ and $\phi_2$ circulate (as one would expect from a non-MMR configuration, thus confirming our previous finding), while $\Delta\omega$ librates in an anti-aligned ($180^\circ$) configuration. Following a more quantitative approach, we repeated the same analysis on $10\,000$ random samples from the posterior and found that for 100\% of them $\Delta\omega$ is confined between $\sim 140^\circ$ and $\sim 220^\circ$ with a mean value perfectly centered on $180^\circ$, therefore confirming the anti-aligned scenario.} 

\subsection{Candidate planet K2-24d}
It is worth noting that we independently confirmed the RV signal of the planet candidate K2-24d -- previously only tentatively detected by \citetalias{Petigura2018} on HIRES data alone. While its parameters appear consistent, our circular fit yielded an 8-$\sigma$ detection at $M_d/M_\oplus = 54_{-4}^{+9}$. The period ratio with respect to the inner planets is too large (and too far from an MMR) to generate a detectable TTV on the inner planets; hence all the constraints on $M_d$ comes from RVs. At $P_d\simeq 470$~d, the a priori transit probability \citep{Winn2010} of planet d would be approximately just $R_\star/a \simeq 0.4\%$, yet the actual chances are much better than that since multiple planetary systems are very likely to be coplanar \citep{Fabrycky2014}. Unfortunately, the fraction of orbital phase currently mapped by K2, Spitzer, and CHEOPS together (all of which are capable of detecting the transit of a $\sim$50~$M_\oplus$ planet at high confidence) is only $<20\%$, so no conclusion can be drawn about the orbital inclination of -d. We will continue to consider planet d as a candidate rather than a confirmed planet since we did not run any specific validation test for it, as such tests are outside the main scope of this paper.

\subsection{Future prospects for follow-up}
The K2-24 system appears to be a very promising target for a follow-up with several current and future facilities. To this purpose, the list of predicted transit windows we
tabulated (available in electronic form at CDS)
is crucial to reliably schedule observations. The most obvious science case is a deeper study of its dynamical architecture, including modeling of new transit timings that could unveil additional companions on orbits on external MMRs to planet c.\footnote{A planet internal to -b (i.e., at $P\lesssim 20$~d) is easily discarded by K2 photometry if it is on transiting configurations. Even if we postulate a non-transiting geometry due to an unusually high mutual inclination, the currently available RVs would put an upper limit to its mass in the rocky planet regime.} Both planets, and -c in particular, are also compelling targets for transmission spectroscopy since their low bulk density combined with the brightness of their host star ($V\simeq 11.3$, $J\simeq 9.6$, $K\simeq 9.2$) offers a unique opportunity to probe the atmospheres of a pair of warm sub-Saturns close to an MMR and to link their composition with their formation site and migration history \citep{Libby2020}. If we compute the transmission spectroscopy metric (TSM; \citealt{Kempton2018}) based on our newly derived parameters on Tables~\ref{tab:stellar_param}, \ref{tab:pyorbit} and \ref{tab:trades}, we get $62\pm 5$ for -b and $177\pm 16$ for -c (TSM-scaled factor computer for $>4$~$R_\oplus$ planets). We highlight that a value of 90 is usually considered the threshold to select the best targets amenable to atmospheric characterization with JWST \citep{Kempton2018}. As already mentioned (Section~\ref{sec:observations_hst}), HST, through WFC3/NIR, has already been exploited to search for atmospheric features on K2-24b, unfortunately with a null result. \citet{Edwards2023b} noted, however, that the best-fit free chemistry model was preferred to a flat line at 2.5$\sigma$, suggesting the presence of NH$_3$, but without evidence for H$_2$O. 

For the first time, TESS will observe K2-24 in Sector 91 of Cycle 7, currently planned from 2025 April 9 to May 7. According to our modeling, only one event will be captured: a transit of K2-24b at 2025-04-25T21:15:40 UTC, unfortunately close to the mid-sector gap. It is difficult at this stage to predict whether or not TESS will manage to add new data to the TTV analysis.
The availability of a new, well-constrained ephemeris, on the other hand, opens an interesting opportunity for a ground-based follow-up campaign from the southern hemisphere. Both transit depths (approx.~$2\,000$ and $4\,000$~ppm, respectively) are feasible, with most medium-sized telescopes operating with the defocusing technique \citep{Nascimbeni2011}, and even partial transits would provide reliable transit times and help in mapping the TTV signal, as the transit shape parameters of both planets (including duration) are now constrained at high precision (Table~\ref{tab:pyorbit}).

As a closing note, we mention that in the next years both PLATO \citep{Rauer2014} and Ariel \citep{Tinetti2021} could provide follow-up of K2-24. PLATO, to be launched in 2026, will unfortunately not observe this target during its long-pointing operation phase since K2-24 lies too close to the Ecliptic to meet the engineering constraints. However, it could be monitored at a later stage for a shorter duration (two to three months) during the so-called short-duration observing phase (\citealt{Nascimbeni2022}).
Ariel, on the other hand, will observe transits of K2-24b and -c in Tier 1 and 3, respectively \citep{Edwards2022}. A detailed study \citep{Borsato2022} demonstrated that the Ariel FGS light curves of K2-24 can also be exploited for accurate TTV analysis and that ten transits would be enough to constrain the presence of an external resonant companion down to the rocky regime. 

\begin{acknowledgements}

We thank the anonymous referee for her/his valuable comments and suggestions.
CHEOPS is an ESA mission in partnership with Switzerland with important contributions to the payload and the ground segment from Austria, Belgium, France, Germany, Hungary, Italy, Portugal, Spain, Sweden, and the United Kingdom. The CHEOPS Consortium would like to gratefully acknowledge the support received by all the agencies, offices, universities, and industries involved. Their flexibility and willingness to explore new approaches were essential to the success of this mission. CHEOPS data analyzed in this article will be made available in the CHEOPS mission archive (\url{https://cheops.unige.ch/archive_browser/}). 
VNa, LBo, TZi, GPi, GMa, IPa, RRa, and GSc acknowledge support from CHEOPS ASI-INAF agreement n. 2019-29-HH.0. 
S.G.S. acknowledge support from FCT through FCT contract nr. CEECIND/00826/2018 and POPH/FSE (EC). 
The Portuguese team thanks the Portuguese Space Agency for the provision of financial support in the framework of the PRODEX Programme of the European Space Agency (ESA) under contract number 4000142255. 
TWi acknowledges support from the UKSA and the University of Warwick. 
YAl acknowledges support from the Swiss National Science Foundation (SNSF) under grant 200020\_192038. 
We acknowledge financial support from the Agencia Estatal de Investigación of the Ministerio de Ciencia e Innovación MCIN/AEI/10.13039/501100011033 and the ERDF “A way of making Europe” through projects PID2019-107061GB-C61, PID2019-107061GB-C66, PID2021-125627OB-C31, and PID2021-125627OB-C32, from the Centre of Excellence “Severo Ochoa” award to the Instituto de Astrofísica de Canarias (CEX2019-000920-S), from the Centre of Excellence “María de Maeztu” award to the Institut de Ciències de l’Espai (CEX2020-001058-M), and from the Generalitat de Catalunya/CERCA programme. 
DBa, EPa, and IRi acknowledge financial support from the Agencia Estatal de Investigación of the Ministerio de Ciencia e Innovación MCIN/AEI/10.13039/501100011033 and the ERDF “A way of making Europe” through projects PID2019-107061GB-C61, PID2019-107061GB-C66, PID2021-125627OB-C31, and PID2021-125627OB-C32, from the Centre of Excellence “Severo Ochoa'' award to the Instituto de Astrofísica de Canarias (CEX2019-000920-S), from the Centre of Excellence “María de Maeztu” award to the Institut de Ciències de l’Espai (CEX2020-001058-M), and from the Generalitat de Catalunya/CERCA programme. A.C.M.C. acknowledges support from the FCT, Portugal, through the CFisUC projects UIDB/04564/2020 and UIDP/04564/2020, with DOI identifiers 10.54499/UIDB/04564/2020 and 10.54499/UIDP/04564/2020, respectively. 
S.C.C.B. acknowledges support from FCT through FCT contracts nr. IF/01312/2014/CP1215/CT0004. 
0000-0003-0312-313X. 
A.C., A.D., B.E., K.G., and J.K. acknowledge their role as ESA-appointed CHEOPS Science Team Members.
ABr was supported by the SNSA. 
CBr and ASi acknowledge support from the Swiss Space Office through the ESA PRODEX program. 
ACC acknowledges support from STFC consolidated grant number ST/V000861/1, and UKSA grant number ST/X002217/1. 
P.E.C. is funded by the Austrian Science Fund (FWF) Erwin Schroedinger Fellowship, program J4595-N. 
This project was supported by the CNES. 
The Belgian participation to CHEOPS has been supported by the Belgian Federal Science Policy Office (BELSPO) in the framework of the PRODEX Program, and by the University of Liège through an ARC grant for Concerted Research Actions financed by the Wallonia-Brussels Federation. 
L.D. thanks the Belgian Federal Science Policy Office (BELSPO) for the provision of financial support in the framework of the PRODEX Programme of the European Space Agency (ESA) under contract number 4000142531. 
This work was supported by FCT - Funda\c{c}\~{a}o para a Ci\^{e}ncia e a Tecnologia through national funds and by FEDER through COMPETE2020 through the research grants UIDB/04434/2020, UIDP/04434/2020, 2022.06962.PTDC. 
O.D.S.D. is supported in the form of work contract (DL 57/2016/CP1364/CT0004) funded by national funds through FCT. 
B.-O. D. acknowledges support from the Swiss State Secretariat for Education, Research and Innovation (SERI) under contract number MB22.00046. 
This project has received funding from the Swiss National Science Foundation for project 200021\_200726. It has also been carried out within the framework of the National Centre of Competence in Research PlanetS supported by the Swiss National Science Foundation under grant 51NF40\_205606. The authors acknowledge the financial support of the SNSF. 
MF and CMP gratefully acknowledge the support of the Swedish National Space Agency (DNR 65/19, 174/18). 
DG gratefully acknowledges financial support from the CRT foundation under Grant No. 2018.2323 “Gaseous or rocky? Unveiling the nature of small worlds”. 
M.G. is an F.R.S.-FNRS Senior Research Associate. 
MNG is the ESA CHEOPS Project Scientist and Mission Representative, and as such also responsible for the Guest Observers (GO) Programme. MNG does not relay proprietary information between the GO and Guaranteed Time Observation (GTO) Programmes, and does not decide on the definition and target selection of the GTO Programme. 
CHe acknowledges support from the European Union H2020-MSCA-ITN-2019 under Grant Agreement no. 860470 (CHAMELEON). 
KGI is the ESA CHEOPS Project Scientist and is responsible for the ESA CHEOPS Guest Observers Programme. She does not participate in, or contribute to, the definition of the Guaranteed Time Programme of the CHEOPS mission through which observations described in this paper have been taken, nor to any aspect of target selection for the programme. 
K.W.F.L. was supported by Deutsche Forschungsgemeinschaft grants RA714/14-1 within the DFG Schwerpunkt SPP 1992, Exploring the Diversity of Extrasolar Planets. 
This work was granted access to the HPC resources of MesoPSL financed by the Region Ile de France and the project Equip@Meso (reference ANR-10-EQPX-29-01) of the programme Investissements d'Avenir supervised by the Agence Nationale pour la Recherche. 
ML acknowledges support of the Swiss National Science Foundation under grant number PCEFP2\_194576. 
PM acknowledges support from STFC research grant number ST/R000638/1. 
This work was also partially supported by a grant from the Simons Foundation (PI Queloz, grant number 327127). 
NCSa acknowledges funding by the European Union (ERC, FIERCE, 101052347). Views and opinions expressed are however those of the author(s) only and do not necessarily reflect those of the European Union or the European Research Council. Neither the European Union nor the granting authority can be held responsible for them. 
GyMSz acknowledges the support of the Hungarian National Research, Development and Innovation Office (NKFIH) grant K-125015, a a PRODEX Experiment Agreement No. 4000137122, the Lend\"ulet LP2018-7/2021 grant of the Hungarian Academy of Science and the support of the city of Szombathely. 
V.V.G. is an F.R.S-FNRS Research Associate. 
JV acknowledges support from the Swiss National Science Foundation (SNSF) under grant PZ00P2\_208945. 
NAW acknowledges UKSA grant ST/R004838/1.
Ple acknowledges that this publication was produced while attending the PhD program in Space Science and Technology at the University of Trento, Cycle XXXVIII, with the support of a scholarship co-financed by the Ministerial Decree no. 351 of 9th April 2022, based on the NRRP - funded by the European Union - NextGenerationEU - Mission 4 "Education and Research", Component 2 "From Research to Business", Investment 3.3 -- CUP E63C22001340001.
E.V. acknowledges support from the ’DISCOBOLO’ project funded by the Spanish Ministerio de Ciencia, Innovación y Universidades under grant PID2021-127289NB-I00.
This work has been carried out within the framework of the NCCR PlanetS supported by the Swiss National Science Foundation under grants 51NF40\_182901 and 51NF40\_205606. AL acknowledges support of the Swiss National Science Foundation under grant number  TMSGI2\_211697.

\end{acknowledgements}

\bibliographystyle{aa}
\bibliography{biblio}

\begin{appendix}

\clearpage

\onecolumn

\section{Additional plots and tables}

\begin{figure*}[!h]
    \centering
    \includegraphics[width=0.45\textwidth, trim=0 0 0 0, clip]{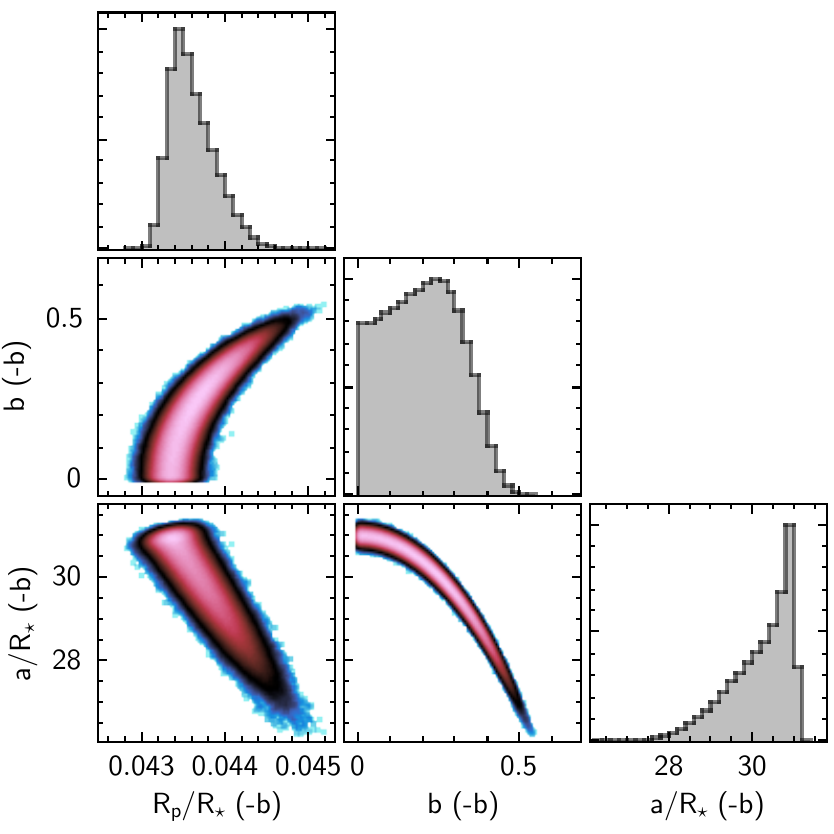}\hspace{5mm}\includegraphics[width=0.45\textwidth, trim=0 0 0 0, clip]{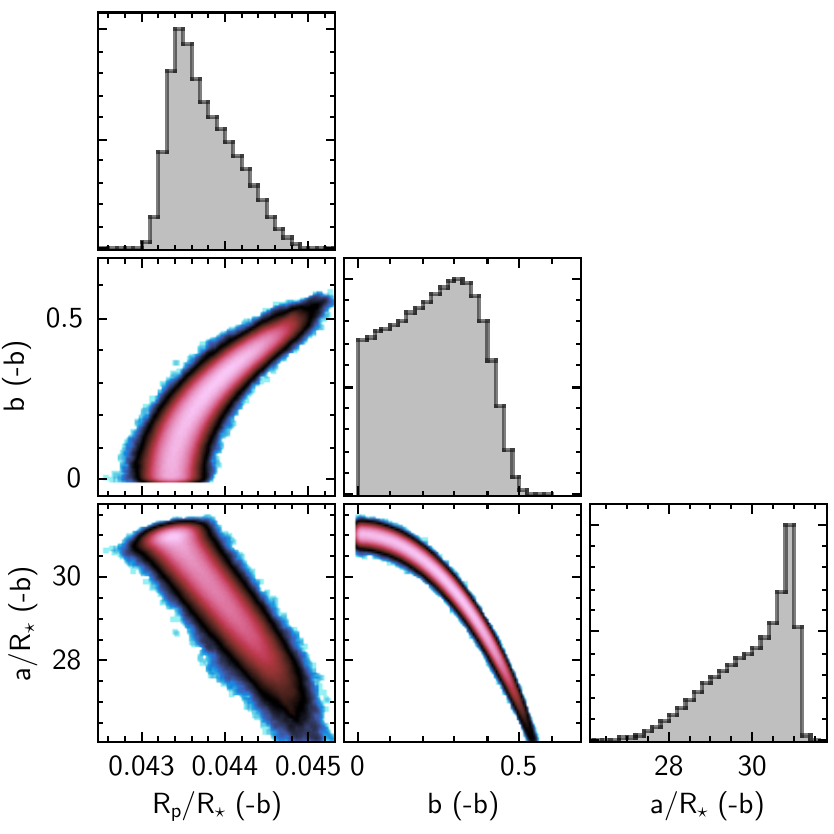}
    \caption{Corner plot of the final MCMC distribution for the transit parameters of planet K2-24b (radius ratio $R_p/R_\star$, impact parameter $b$, scaled semi-major axis $a/R_\star$) of our LD-prior (left panel) and LD-fit (right panel) photometric modeling described in Section~\ref{sec:photometric}. The best-fit values are tabulated in Table~\ref{tab:pyorbit}, second and fourth column, respectively.}
    \label{fig:corner_b}
\end{figure*}

\begin{figure*}[!h]
    \centering
    \includegraphics[width=0.45\textwidth, trim=0 0 0 0, clip]{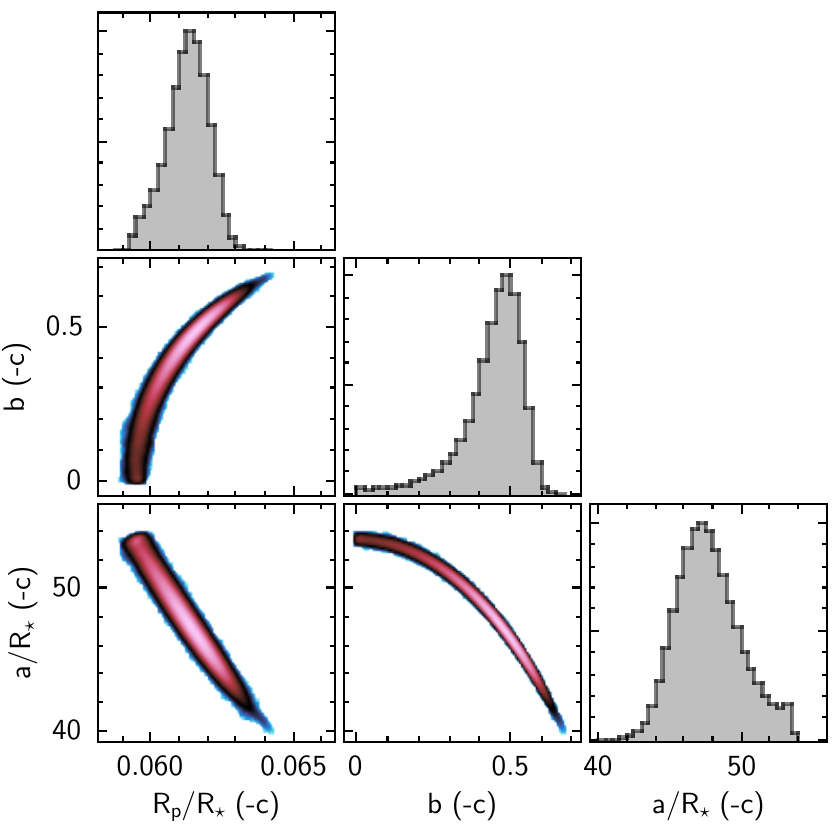}\hspace{5mm}\includegraphics[width=0.45\textwidth, trim=0 0 0 0, clip]{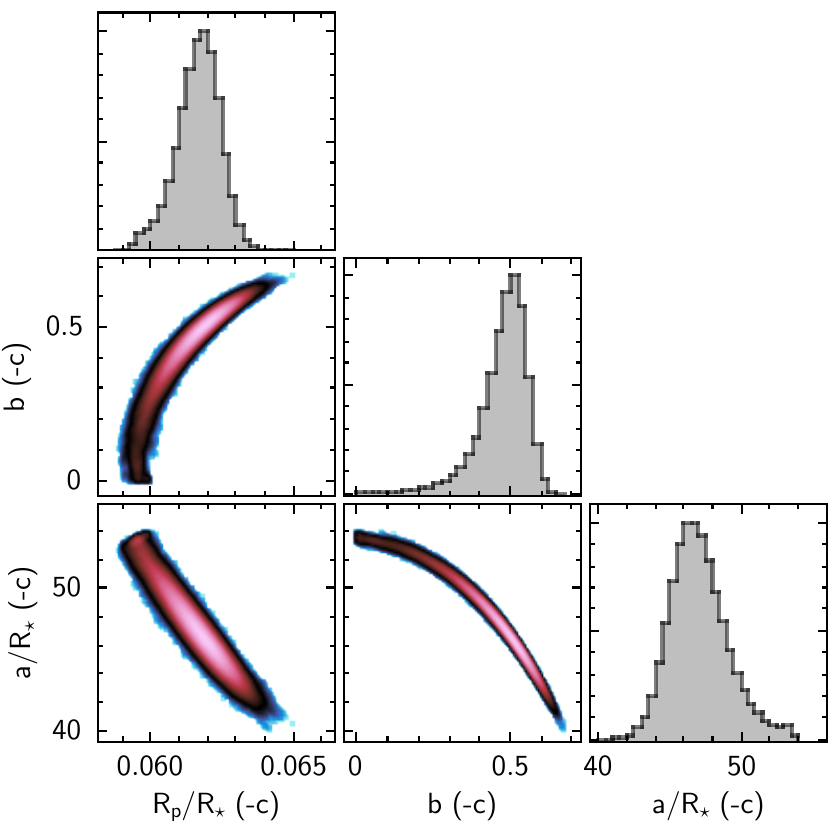}
    \caption{Corner plot of the final MCMC distribution for the transit parameters of planet K2-24c (radius ratio $R_p/R_\star$, impact parameter $b$, scaled semi-major axis $a/R_\star$) of our LD-prior (left panel) and LD-fit (right panel) photometric modeling described in Section~\ref{sec:photometric}. The best-fit values are tabulated in Table~\ref{tab:pyorbit}, second and fourth column, respectively.}
    \label{fig:corner_c}
\end{figure*}

\begin{table*}
    \centering\centering\small\renewcommand{\arraystretch}{1.2}
        \caption{Log of the CHEOPS observations.}
    \begin{tabular}{cccccc} \hline \hline
    planet & ID & UT date & $N$ & $\sigma$ [ppm] & key \\ \hline
K2-24b & \texttt{b1} & 2021-06-26 & 392 & 606 &\texttt{CH\_PR100025\_TG006001\_V0300} \\	
K2-24b & \texttt{b2} & 2021-07-16 & 359 & 642 &\texttt{CH\_PR100025\_TG006002\_V0300} \\	
K2-24c & \texttt{c1} & 2021-07-17 & 433 & 590 &\texttt{CH\_PR100025\_TG006101\_V0300} \\	
K2-24c & \texttt{c2} & 2022-03-28 & 480 & 685 &\texttt{CH\_PR100025\_TG007101\_V0300} \\	
K2-24b & \texttt{b3} & 2022-05-05 & 427 & 732 &\texttt{CH\_PR100025\_TG007001\_V0300} \\	
K2-24c & \texttt{c3} & 2022-05-10 & 563 & 735 &\texttt{CH\_PR100025\_TG007102\_V0300} \\	
K2-24b & \texttt{b4} & 2022-05-26 & 433 & 671 &\texttt{CH\_PR100025\_TG007002\_V0300} \\	
K2-24b & \texttt{b5} & 2022-06-16 & 368 & 769 &\texttt{CH\_PR100025\_TG007003\_V0300} \\	
K2-24c & \texttt{c4} & 2022-06-21 & 468 & 728 &\texttt{CH\_PR100025\_TG007103\_V0300} \\	
K2-24c & \texttt{c5} & 2023-04-13 & 496 & 677 &\texttt{CH\_PR100025\_TG007401\_V0300} \\	
K2-24c & \texttt{c6} & 2023-05-26 & 504 & 647 &\texttt{CH\_PR100025\_TG007901\_V0300} \\	
K2-24b & \texttt{b6} & 2023-06-06 & 576 & 672 &\texttt{CH\_PR100025\_TG008001\_V0300} \\	
K2-24b & \texttt{b7} & 2023-06-27 & 490 & 618 &\texttt{CH\_PR100025\_TG008101\_V0300} \\ \hline
    \end{tabular}\tablefoot{The columns give: the planet name (K2-24b or -c), the ID matching the labels in Fig.~\ref{fig:lc}, the UT date at start, the number of photometric points $N$, the root mean square $\sigma$ with respect to the best-fit model in parts per million,  and the CHEOPS key ID.}
    \label{tab:log}
\end{table*}

\begin{figure*}
    \centering
    \includegraphics[width=0.45\columnwidth, trim=0 0 0 0]{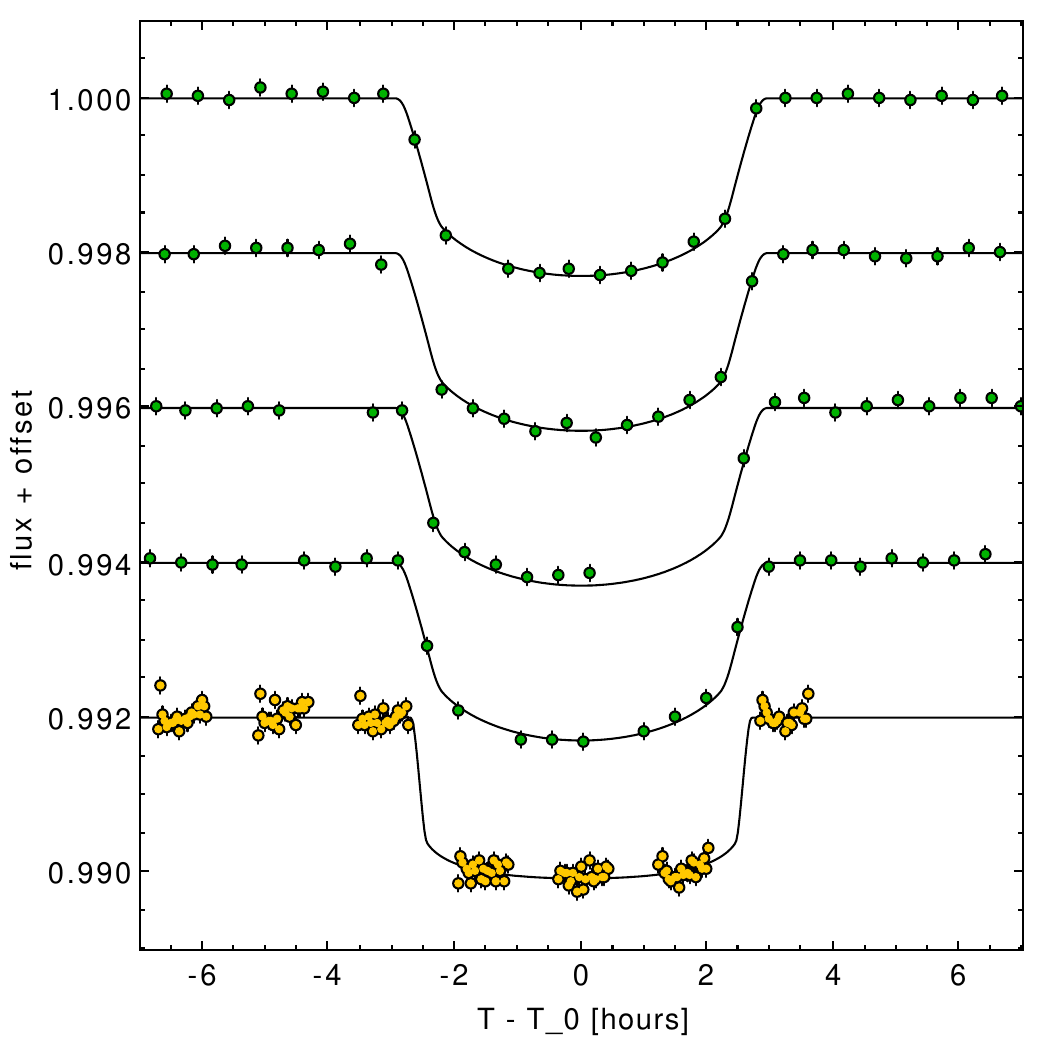}\hspace{3mm}\includegraphics[width=0.45\columnwidth, trim=0 0 0 0]{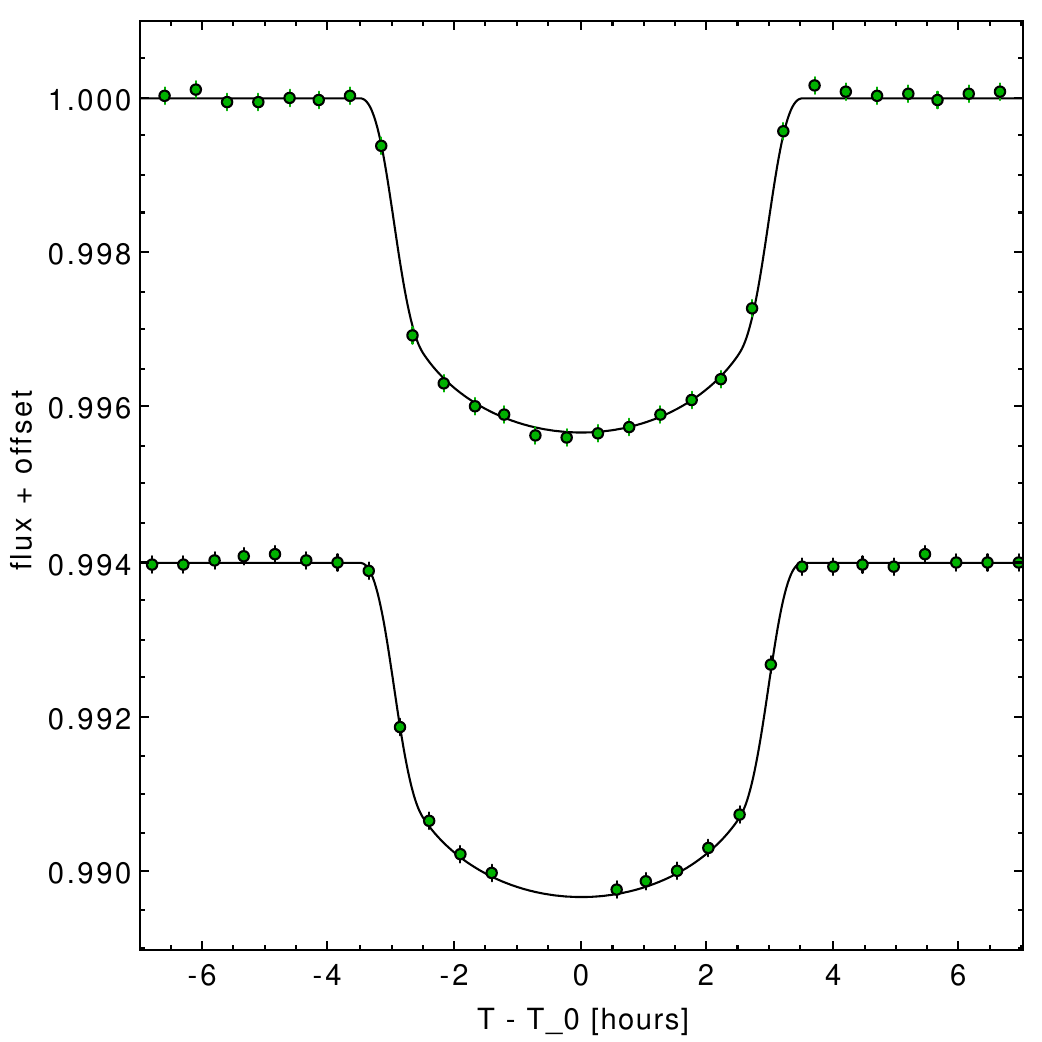}
\caption{Light curves of K2-24b (left panel) and K2-24c (right panel) from K2 (green circles) and HST/WFC3 (yellow circles) analyzed for the present work (Section~\ref{sec:observations_k2} and \ref{sec:observations_hst}). The light curves are sorted in chronological order from top to bottom, as in Table~\ref{tab:t0}. An arbitrary vertical offset was added for visualization.}
    \label{fig:lck2hst}
\end{figure*}

\begin{figure*}
    \centering
    \includegraphics[width=0.9\columnwidth, trim=0 0 0 0]{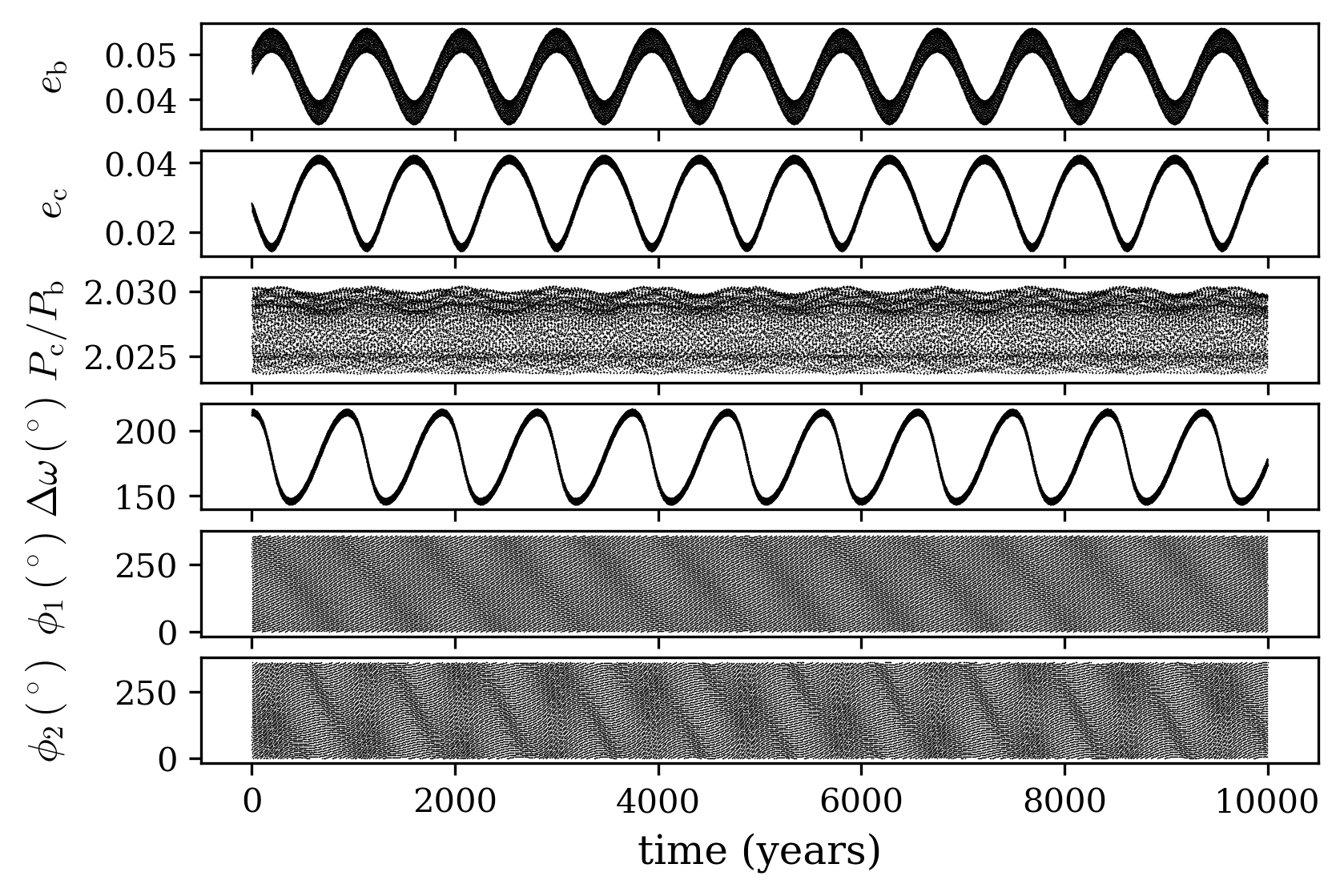}
\caption{\valerio{Evolution of the orbital parameters of K2-24b and -c, integrated over $10\,000$ years (see Section~\ref{sec:mmr} for details). From top to bottom: eccentricities $e_b$, $e_c$ of planet -b and -c (respectively), orbital period ratio $P_c/P_b$, difference between the pericenter argument $\Delta\omega$, critical resonant angles $\phi_1$ and $\phi_2$.}}
    \label{fig:evolution}
\end{figure*}

\FloatBarrier

\begin{table}
    \centering\centering\small\renewcommand{\arraystretch}{1.2}
        \caption{Transit times of K2-24b and K2-24c predicted by our dynamic model for 2024-2025.}
    \begin{tabular}{ccccc} \hline \hline
Planet & $T_0$~[JD] & $T_0$ [$\textrm{BJD}_\textrm{TDB}$] & $\sigma(T_0)$~[d] & $T_0$~[UTC]\\ \hline
  -b  &   2460310.7396 & 2460310.7434& 0.0013 & 2024-01-01T05:50:34 \\
  -b  &   2460331.6350 & 2460331.6373& 0.0016 & 2024-01-22T03:17:47 \\
  -c  &   2460345.4217 & 2460345.4228& 0.0021 & 2024-02-04T22:08:52 \\
  -b  &   2460352.5281 & 2460352.5285& 0.0016 & 2024-02-12T00:41:08 \\
  -b  &   2460373.4251 & 2460373.4235& 0.0019 & 2024-03-03T22:09:52 \\
  -c  &   2460387.7410 & 2460387.7379& 0.0023 & 2024-03-18T05:42:40 \\
  -b  &   2460394.3197 & 2460394.3161& 0.0020  & 2024-03-24T19:35:12 \\
  -b  &   2460415.2182 & 2460415.2130& 0.0022 & 2024-04-14T17:06:43 \\
  -c  &   2460430.0539 & 2460430.0478& 0.0027 & 2024-04-29T13:08:54 \\
  -b  &   2460436.1141 & 2460436.1078& 0.0022 & 2024-05-05T14:35:18 \\
  -b  &   2460457.0139 & 2460457.0072& 0.0024 & 2024-05-26T12:10:25 \\
  -c  &   2460472.3612 & 2460472.3547& 0.0029 & 2024-06-10T20:30:48 \\
  -b  &   2460477.9110 & 2460477.9046& 0.0025 & 2024-06-16T09:42:43 \\
  -b  &   2460498.8116 & 2460498.8063& 0.0026 & 2024-07-07T07:21:09 \\
  -c  &   2460514.6640 & 2460514.6598& 0.0029 & 2024-07-23T03:50:14 \\
  -b  &   2460519.7097 & 2460519.7059& 0.0026 & 2024-07-28T04:56:37 \\
  -b  &   2460540.6109 & 2460540.6090& 0.0027 & 2024-08-18T02:37:04 \\
  -c  &   2460556.9639 & 2460556.9636& 0.0030 & 2024-09-03T11:07:42 \\
  -b  &   2460561.5096 & 2460561.5098& 0.0027 & 2024-09-08T00:14:12 \\
  -b  &   2460582.4109 & 2460582.4130& 0.0027 & 2024-09-28T21:54:43 \\
  -c  &   2460599.2626 & 2460599.2659& 0.0031 & 2024-10-15T18:23:00 \\
  -b  &   2460603.3099 & 2460603.3135& 0.0026 & 2024-10-19T19:31:33 \\
  -b  &   2460624.2107 & 2460624.2153& 0.0026 & 2024-11-09T17:10:03 \\
  -c  &   2460641.5619 & 2460641.5667& 0.0030 & 2024-11-27T01:36:11 \\
  -b  &   2460645.1096 & 2460645.1144& 0.0025 & 2024-11-30T14:44:51 \\
  -b  &   2460666.0095 & 2460666.0138& 0.0024 & 2024-12-21T12:19:55 \\
  -c  &   2460683.8638 & 2460683.8671& 0.0028 & 2025-01-08T08:48:45 \\
  -b  &   2460686.9079 & 2460686.9110& 0.0023 & 2025-01-11T09:51:51 \\
  -b  &   2460707.8064 & 2460707.8077& 0.0022 & 2025-02-01T07:23:13 \\
  -c  &   2460726.1700 & 2460726.1696& 0.0027 & 2025-02-19T16:04:14 \\
  -b  &   2460728.7038 & 2460728.7031& 0.0020 & 2025-02-22T04:52:35 \\
  -b  &   2460749.6007 & 2460749.5979& 0.0019 & 2025-03-15T02:21:05 \\
  -c  &   2460768.4818 & 2460768.4774& 0.0025 & 2025-04-02T23:27:29 \\
  -b  &   2460770.4968 & 2460770.4922& 0.0018 & 2025-04-04T23:48:54 \\
  -b  &   2460791.3917 & 2460791.3858& 0.0016 & 2025-04-25T21:15:40 \\
  -c  &   2460810.8002 & 2460810.7937& 0.0022 & 2025-05-15T07:02:56 \\
  -b  &   2460812.2863 & 2460812.2797& 0.0016 & 2025-05-16T18:42:48 \\
  -b  &   2460833.1792 & 2460833.1726& 0.0015 & 2025-06-06T16:08:35 \\
  -c  &   2460853.1260 & 2460853.1201& 0.0020 & 2025-06-26T14:53:01 \\
  -b  &   2460854.0720 & 2460854.0661& 0.0014 & 2025-06-27T13:35:13 \\
  -b  &   2460874.9628 & 2460874.9583& 0.0013 & 2025-07-18T10:59:58 \\
  -c  &   2460895.4595 & 2460895.4566& 0.0019 & 2025-08-07T22:57:37 \\
  -b  &   2460895.8538 & 2460895.8510& 0.0013 & 2025-08-08T08:25:29 \\
  -b  &   2460916.7427 & 2460916.7419& 0.0012 & 2025-08-29T05:48:25 \\
  -b  &   2460937.6318 & 2460937.6330& 0.0012 & 2025-09-19T03:11:37 \\
  -c  &   2460937.8003 & 2460937.8016& 0.0019 & 2025-09-19T07:14:18 \\
  -b  &   2460958.5189 & 2460958.5219& 0.0012 & 2025-10-10T00:31:33 \\
  -b  &   2460979.4063 & 2460979.4105& 0.0012 & 2025-10-30T21:51:10 \\
  -c  &   2460980.1481 & 2460980.1524& 0.0019 & 2025-10-31T15:39:29 \\
  -b  &   2461000.2918 & 2461000.2966& 0.0011 & 2025-11-20T19:07:09 \\
  -b  &   2461021.1776 & 2461021.1822& 0.0011 & 2025-12-11T16:22:29 \\
  -c  &   2461022.5022 & 2461022.5068& 0.0018 & 2025-12-13T00:09:54 \\ \hline
    \end{tabular}\tablefoot{The columns give: the planet name (K2-24, -b or -c), the transit time $T_0$ predicted by our best-fit dynamical solution (see Section~\ref{sec:predictions} for details) in the JD and BJD-TDB standard along with its 1-$\sigma$ error bar (symmetrized), and the $T_0$ in the UTC-ISO standard.}
    \label{tab:t0_pred1}
\end{table}

\begin{table}
    \centering\centering\small\renewcommand{\arraystretch}{1.2}
        \caption{Transit times of K2-24b and K2-24c predicted by our dynamic model for 2026-2027.}
    \begin{tabular}{ccccc} \hline \hline
Planet & $T_0$~[JD] & $T_0$ [$\textrm{BJD}_\textrm{TDB}$] & $\sigma(T_0)$~[d] & $T_0$~[UTC]\\ \hline
  -b  &   2461042.0617 & 2461042.0655& 0.0011 & 2026-01-01T13:34:22 \\
  -b  &   2461062.9461 & 2461062.9484& 0.0011 & 2026-01-22T10:45:46 \\
  -c  &   2461064.8618 & 2461064.8639& 0.0017 & 2026-01-24T08:44:05 \\
  -b  &   2461083.8292 & 2461083.8295& 0.0011 & 2026-02-12T07:54:33 \\
  -b  &   2461104.7125 & 2461104.7107& 0.0010 & 2026-03-05T05:03:30 \\
  -c  &   2461107.2261 & 2461107.2242& 0.0016 & 2026-03-07T17:22:51 \\
  -b  &   2461125.5946 & 2461125.5909& 0.0010 & 2026-03-26T02:10:57 \\
  -b  &   2461146.4770 & 2461146.4717& 0.0010 & 2026-04-15T23:19:18 \\
  -c  &   2461149.5944 & 2461149.5889& 0.0014 & 2026-04-19T02:08:05 \\
  -b  &   2461167.3585 & 2461167.3522& 0.0010 & 2026-05-06T20:27:10 \\
  -b  &   2461188.2402 & 2461188.2335& 0.0010 & 2026-05-27T17:36:21 \\
  -c  &   2461191.9657 & 2461191.9590& 0.0013 & 2026-05-31T11:01:02 \\
  -b  &   2461209.1213 & 2461209.1150& 0.0010  & 2026-06-17T14:45:38 \\
  -b  &   2461230.0026 & 2461229.9973& 0.0010 & 2026-07-08T11:56:09 \\
  -c  &   2461234.3392 & 2461234.3342& 0.0012 & 2026-07-12T20:01:18 \\
  -b  &   2461250.8834 & 2461250.8797& 0.0010 & 2026-07-29T09:06:53 \\
  -b  &   2461271.7645 & 2461271.7627& 0.0010 & 2026-08-19T06:18:21 \\
  -c  &   2461276.7142 & 2461276.7129& 0.0013 & 2026-08-24T05:06:40 \\
  -b  &   2461292.6453 & 2461292.6456& 0.0010 & 2026-09-09T03:29:40 \\
  -b  &   2461313.5263 & 2461313.5285& 0.0010 & 2026-09-30T00:41:04 \\
  -c  &   2461319.0901 & 2461319.0927& 0.0014 & 2026-10-05T14:13:34 \\
  -b  &   2461334.4073 & 2461334.4109& 0.0010 & 2026-10-20T21:51:49 \\
  -b  &   2461355.2884 & 2461355.2930& 0.0010 & 2026-11-10T19:02:01 \\
  -c  &   2461361.4662 & 2461361.4710& 0.0015 & 2026-11-16T23:18:18 \\
  -b  &   2461376.1696 & 2461376.1744& 0.0010 & 2026-12-01T16:11:15 \\
  -b  &   2461397.0510 & 2461397.0553& 0.0010 & 2026-12-22T13:19:46 \\
  -c  &   2461403.8420 & 2461403.8460& 0.0016 & 2026-12-29T08:18:18 \\
  -b  &   2461417.9326 & 2461417.9357& 0.0010 & 2027-01-12T10:27:25 \\
  -b  &   2461438.8145 & 2461438.8158& 0.0010 & 2027-02-02T07:34:51 \\
  -c  &   2461446.2169 & 2461446.2175& 0.0017 & 2027-02-09T17:13:15 \\
  -b  &   2461459.6965 & 2461459.6958& 0.0010 & 2027-02-23T04:42:00 \\
  -b  &   2461480.5790 & 2461480.5763& 0.0010 & 2027-03-16T01:49:52 \\
  -c  &   2461488.5901 & 2461488.5866& 0.0018 & 2027-03-24T02:04:50 \\
  -b  &   2461501.4616 & 2461501.4570& 0.0010 & 2027-04-05T22:58:10 \\
  -b  &   2461522.3448 & 2461522.3389& 0.0010 & 2027-04-26T20:08:09 \\
  -c  &   2461530.9613 & 2461530.9550& 0.0020 & 2027-05-05T10:55:20 \\
  -b  &   2461543.2281 & 2461543.2215& 0.0010 & 2027-05-17T17:19:00 \\
  -b  &   2461564.1122 & 2461564.1056& 0.0010 & 2027-06-07T14:32:11 \\
  -c  &   2461573.3298 & 2461573.3235& 0.0020 & 2027-06-16T19:45:50 \\
  -b  &   2461584.9962 & 2461584.9903& 0.0010 & 2027-06-28T11:46:08 \\
  -b  &   2461605.8813 & 2461605.8768& 0.0010 & 2027-07-19T09:02:41 \\
  -c  &   2461615.6949 & 2461615.6912& 0.0021 & 2027-07-29T04:35:24 \\
  -b  &   2461626.7662 & 2461626.7634& 0.0010 & 2027-08-09T06:19:22 \\
  -b  &   2461647.6525 & 2461647.6518& 0.0010 & 2027-08-30T03:38:35 \\
  -c  &   2461658.0561 & 2461658.0564& 0.0021 & 2027-09-09T13:21:14 \\
  -b  &   2461668.5382 & 2461668.5395& 0.0010 & 2027-09-20T00:56:57 \\
  -b  &   2461689.4259 & 2461689.4289& 0.0011 & 2027-10-10T22:17:39 \\
  -c  &   2461700.4125 & 2461700.4162& 0.0021 & 2027-10-21T21:59:27 \\
  -b  &   2461710.3127 & 2461710.3169& 0.0011 & 2027-10-31T19:36:25 \\
  -b  &   2461731.2017 & 2461731.2066& 0.0012 & 2027-11-21T16:57:32 \\
  -c  &   2461742.7636 & 2461742.7684& 0.0021 & 2027-12-03T06:26:33 \\
  -b  &   2461752.0897 & 2461752.0944& 0.0012 & 2027-12-12T14:15:57 \\ \hline
    \end{tabular}\tablefoot{The columns give: the planet name (K2-24, -b or -c), the transit time $T_0$ predicted by our best-fit dynamical solution (see Section~\ref{sec:predictions} for details) in the JD and BJD-TDB standard along with its 1-$\sigma$ error bar (symmetrized), and the $T_0$ in the UTC-ISO standard.}
    \label{tab:t0_pred2}
\end{table}

\begin{table}
    \centering\centering\small\renewcommand{\arraystretch}{1.2}
        \caption{Transit times of K2-24b and K2-24c predicted by our dynamic model for 2028-2029.}
    \begin{tabular}{ccccc} \hline \hline
Planet & $T_0$~[JD] & $T_0$ [$\textrm{BJD}_\textrm{TDB}$] & $\sigma(T_0)$~[d] & $T_0$~[UTC]\\ \hline
  -b  &   2461772.9803 & 2461772.9841& 0.0014 & 2028-01-02T11:37:11 \\
  -c  &   2461785.1086 & 2461785.1116& 0.0023 & 2028-01-14T14:40:46 \\
  -b  &   2461793.8696 & 2461793.8718& 0.0014 & 2028-01-23T08:55:28 \\
  -b  &   2461814.7619 & 2461814.7622& 0.0016 & 2028-02-13T06:17:36 \\
  -c  &   2461827.4473 & 2461827.4464& 0.0024 & 2028-02-25T22:42:50 \\
  -b  &   2461835.6524 & 2461835.6506& 0.0016 & 2028-03-05T03:37:00 \\
  -b  &   2461856.5464 & 2461856.5427& 0.0018 & 2028-03-26T01:01:33 \\
  -c  &   2461869.7792 & 2461869.7744& 0.0026 & 2028-04-08T06:35:13 \\
  -b  &   2461877.4384 & 2461877.4331& 0.0019 & 2028-04-15T22:23:42 \\
  -b  &   2461898.3341 & 2461898.3278& 0.0022 & 2028-05-06T19:52:04 \\
  -c  &   2461912.1041 & 2461912.0975& 0.0029 & 2028-05-20T14:20:26 \\
  -b  &   2461919.2275 & 2461919.2208& 0.0023 & 2028-05-27T17:18:04 \\
  -b  &   2461940.1248 & 2461940.1186& 0.0026 & 2028-06-17T14:50:48 \\
  -c  &   2461954.4222 & 2461954.4166& 0.0033 & 2028-07-01T21:59:55 \\
  -b  &   2461961.0197 & 2461961.0144& 0.0027 & 2028-07-08T12:20:52 \\
  -b  &   2461981.9185 & 2461981.9148& 0.0029 & 2028-07-29T09:57:26 \\
  -c  &   2461996.7340 & 2461996.7317& 0.0038 & 2028-08-13T05:33:40 \\
  -b  &   2462002.8146 & 2462002.8129& 0.0030 & 2028-08-19T07:30:41 \\
  -b  &   2462023.7147 & 2462023.7150& 0.0032 & 2028-09-09T05:09:39 \\
  -c  &   2462039.0403 & 2462039.0421& 0.0041 & 2028-09-24T13:00:38 \\
  -b  &   2462044.6120 & 2462044.6142& 0.0033 & 2028-09-30T02:44:32 \\
  -b  &   2462065.5129 & 2462065.5166& 0.0034 & 2028-10-21T00:23:58 \\
  -c  &   2462081.3425 & 2462081.3470& 0.0043 & 2028-11-05T20:19:42 \\
  -b  &   2462086.4111 & 2462086.4158& 0.0034 & 2028-11-10T21:58:46 \\
  -b  &   2462107.3124 & 2462107.3173& 0.0035 & 2028-12-01T19:36:58 \\
  -c  &   2462123.6419 & 2462123.6464& 0.0044 & 2028-12-18T03:30:55 \\
  -b  &   2462128.2113 & 2462128.2156& 0.0035 & 2028-12-22T17:10:30 \\
  -b  &   2462149.1126 & 2462149.1156& 0.0035 & 2029-01-12T14:46:33 \\
  -c  &   2462165.9405 & 2462165.9421& 0.0044 & 2029-01-29T10:36:44 \\
  -b  &   2462170.0116 & 2462170.0129& 0.0034 & 2029-02-02T12:18:42 \\
  -b  &   2462190.9124 & 2462190.9116& 0.0034 & 2029-02-23T09:52:50 \\
  -c  &   2462208.2399 & 2462208.2375& 0.0044 & 2029-03-12T17:42:01 \\
  -b  &   2462211.8113 & 2462211.8085& 0.0033 & 2029-03-16T07:24:17 \\
  -b  &   2462232.7110 & 2462232.7065& 0.0032 & 2029-04-06T04:57:21 \\
  -c  &   2462250.5422 & 2462250.5365& 0.0042 & 2029-04-24T00:52:35 \\
  -b  &   2462253.6093 & 2462253.6034& 0.0031 & 2029-04-27T02:28:59 \\
  -b  &   2462274.5077 & 2462274.5011& 0.0029 & 2029-05-18T00:01:38 \\
  -c  &   2462292.8490 & 2462292.8424& 0.0038 & 2029-06-05T08:13:11 \\
  -b  &   2462295.4049 & 2462295.3984& 0.0028 & 2029-06-07T21:33:46 \\
  -b  &   2462316.3015 & 2462316.2958& 0.0026 & 2029-06-28T19:05:57 \\
  -c  &   2462335.1617 & 2462335.1571& 0.0034 & 2029-07-17T15:46:21 \\
  -b  &   2462337.1975 & 2462337.1930& 0.0025 & 2029-07-19T16:38:00 \\
  -b  &   2462358.0921 & 2462358.0895& 0.0024 & 2029-08-09T14:08:54 \\
  -c  &   2462377.4813 & 2462377.4805& 0.0030 & 2029-08-28T23:32:01 \\
  -b  &   2462378.9864 & 2462378.9858& 0.0023 & 2029-08-30T11:39:35 \\
  -b  &   2462399.8790 & 2462399.8804& 0.0021 & 2029-09-20T09:07:50 \\
  -c  &   2462419.8083 & 2462419.8113& 0.0027 & 2029-10-10T07:28:21 \\
  -b  &   2462420.7715 & 2462420.7746& 0.0020 & 2029-10-11T06:35:31 \\
  -b  &   2462441.6621 & 2462441.6664& 0.0019 & 2029-11-01T03:59:42 \\
  -c  &   2462462.1429 & 2462462.1477& 0.0026 & 2029-11-21T15:32:49 \\
  -b  &   2462462.5528 & 2462462.5576& 0.0017 & 2029-11-22T01:23:03 \\
  -b  &   2462483.4414 & 2462483.4461& 0.0016 & 2029-12-12T22:42:23 \\ \hline
    \end{tabular}\tablefoot{The columns give: the planet name (K2-24, -b or -c), the transit time $T_0$ predicted by our best-fit dynamical solution (see Section~\ref{sec:predictions} for details) in the JD and BJD-TDB standard along with its 1-$\sigma$ error bar (symmetrized), and the $T_0$ in the UTC-ISO standard.}
    \label{tab:t0_pred3}
\end{table}


\end{appendix}
\end{document}